\documentclass[12pt]{article}
\usepackage[T1]{fontenc}
\usepackage[utf8]{inputenc}
\usepackage{mathtools}
\usepackage{amsbsy}
\usepackage{amstext}
\usepackage{stmaryrd}
\usepackage{setspace}
\usepackage{amssymb,amsthm,thmtools}
\newtheorem{thm}{Theorem}
\newtheorem{example}{Example}

\usepackage{graphicx}
\usepackage{url}
\usepackage{color}
\usepackage[ruled,linesnumbered]{algorithm2e}
\usepackage{float}

\usepackage{amsmath}
\usepackage{graphicx,psfrag,epsf}
\usepackage{enumerate}
\usepackage[authoryear,round]{natbib}

\usepackage{hyperref}
\usepackage{enumitem}
\usepackage{fullpage}
\usepackage{authblk}

\begin{document}

\def\spacingset#1{\renewcommand{\baselinestretch}
{#1}\small\normalsize} \spacingset{1}

%%%%%%%%%%%%%%%%%%%%%%%%%%%%%%%%%%%%%%%%%%%%%%%%%%%%%%%%%%%%%%%%%%%%%%%%%%%%%%

\title{A Distance-preserving Matrix Sketch}
\author[1, 2]{\small Leland Wilkinson}
\author[3]{\small Hengrui Luo}

\affil[1]{\footnotesize H2O.ai, 2307 Leghorn St, Mountain View, CA 94043, USA, E-mail: leland.wilkinson@h2o.ai}
\affil[2]{\footnotesize Department of Computer Science, University of Illinois at Chicago, }
\affil[3]{\footnotesize Lawrence Berkeley National Laboratory, Berkeley, CA, 94720, USA, E-mail: hrluo@lbl.gov}
\maketitle

\abstract{
Visualizing very large matrices involves many formidable problems. Various popular solutions to these problems involve sampling, clustering, projection, or feature selection to reduce the size and complexity of the original task. An important aspect of these methods is how to preserve relative distances between points in the higher-dimensional space after reducing rows and columns to fit in a lower dimensional space. This aspect is important because conclusions based on faulty visual reasoning can be harmful. Judging dissimilar points as similar or similar points as dissimilar on the basis of a visualization can lead to false conclusions. To ameliorate this bias and to make visualizations of very large datasets feasible, we introduce two new algorithms that respectively select a subset of rows and columns of a rectangular matrix. This selection is designed to preserve relative distances as closely as possible. We compare our matrix sketch to more traditional alternatives on a variety of artificial and real datasets.
} 

%\keywords{Squashing, Aggregation, Big Data, Feature Selection, Matrix Sketch.}
\noindent%

%\spacingset{1.45}
%\doublespacing
\singlespacing

\section{Introduction}

For a real data matrix $X_{np}$ of size $n\times p$, we assume that rows represent points and columns represent dimensions in a real metric space ($\mathbb{R}^p$). We might be interested in visually identifying such features as outliers, duplicate points, anomalies, or unusual distributions. When $n$ and $p$ are moderate in size, we can still use simple plots and statistics to explore such features. When these parameters are larger, however, these simple tasks become unwieldy, as explained below. While it is not uncommon to see machine learning models with $n > 10^9$ and $p > 10^5$, exploratory visualization of datasets smaller than these extremes can be problematic for the following reasons:

\begin{itemize}
\itemsep0em 
\item The data won't fit in memory. We can use a columnar distributed database, but this usually fails to deliver the response times users expect in an interactive exploratory environment \citep{BatchInteractive}
\item Most EDA algorithms do not scale to problems this size \citep{Pixnostics}
\item We cannot send big data "over the wire" in client-server environments where response time is important. 
\item Sampling tends to conceal outliers and other singular features.
\item Plotting many points on display devices (even megapixel or 4K) produces a big opaque spot. We can use kernels, alpha-channel rendering, binning, and other methods to mitigate overlaps, but this impedes brushing and linking gestures.
\item Projections or embedding often violate metric axioms -- points close together in higher-dimensional space may be far apart in lower-dimensional projections. Conversely, points far apart in higher-dimensional space may be close together in a projection \citep{HLuoGeneralized2020}. 
\item Thousands of dimensions overwhelm multivariate displays such as parallel coordinates and scatterplot matrices. They run out of display ``real estate.''
\end{itemize}

\subsection{Our Contribution}
We address these challenges with a pair of algorithms that subset data matrices. We select a subset of rows and columns of $X_{np}$ :\\
\begin{equation*}
X_{np} \mapsto X_{mp} \mapsto X[a,b]_{mk}, 
\end{equation*}
where $m \ll n$ and $k \ll p$ and $a$ is a row index array of length $m$ and $b$ is a column index array of length $k$. We restrict our selection of $X_{mk}$ to be approximately \textit{distance-preserving}, where distances between the rows of $X_{mk}$ are linearly related to the  distances between the corresponding rows of $X_{mp}$. 

In sketching the rows in $X_{np}$, we collect points that are relatively close to each other. Our method depends on computing $m$ Euclidean balls of radius $r$ in $p$-dimensional space. We choose $r$ to be as small as possible when reducing $n$ to a manageable-sized $m$. 

In subsequently sketching columns in $X_{mp}$, we produce a submatrix of $X_{mp}$ based on $k$ of its columns. We select these $k$ columns such that distances between the $m$ rows in $X_{mk}$ linearly approximate the distances between the $m$ rows in $X_{mp}$. Our hope is that this new submatrix $X_{mk}$ is substitutable for $X_{np}$ in visual analytic explorations. Although this is a lossy compression, we are able to retain pointers to the rows and columns of $X_{np}$ that are not in $X_{mk}$. Consequently, we can employ our sketching algorithm as a preprocessor for interactive applications.

Using these two algorithms allows subsequent analysis of $X_{mk}$ based on a representative subset of its original columns using frequency-weighted statistical models; the weights for each row of $X_{mk}$ are derived from the number of points inside each of the $m$ balls. The algorithms are designed to work separately or together. RowSketcher can be used alone on deep matrices (many rows) and ColSketcher can be used alone on wide matrices (many columns). Both can be used successively on large, approximately square matrices. If both RowSketcher and ColSketcher are used on a given dataset, our convention is to sketch whichever dimension is larger ($n$ or $p$) first. This approach improves performance.

\begin{figure}[h]
\centering
\includegraphics[width=12cm]{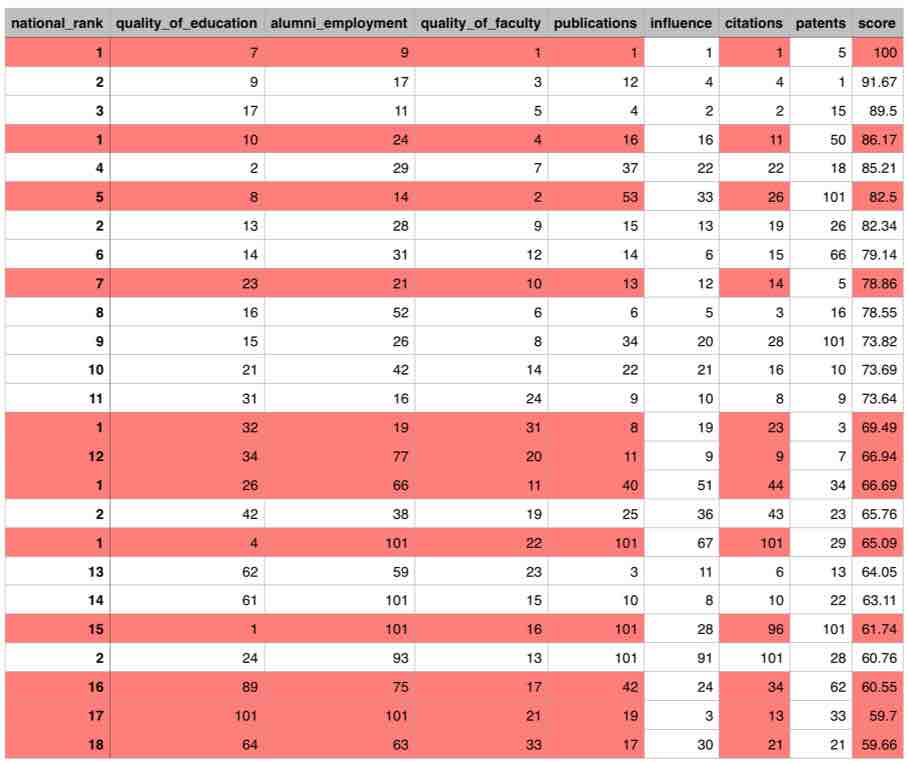}
\caption {Matrix of university ratings.}
\label{fig:universities}
\end{figure}

Figure~\ref{fig:universities} contains data on world universities from the cwurData.csv dataset at \url{https://www.kaggle.com/mylesoneill/world-university-rankings}. The table has been anonymized and consists of a subset of the original dataset. The red cells denote values retained by the row and column sketch algorithms, while the white cells are omitted. The figure shows an actual application of a sketching algorithm so that readers can see exactly what it does.

From the construction side, our algorithms have several distinctive aspects. First, they are scalable to larger datasets with moderate growth of complexity. Second, they work on streaming data and are parallelizable, since updating rows and columns involves additive updates of distances. Third, our algorithms produce axis-parallel results suitable for visualization; it does not create composites of the input columns and retains interpretability of dimensions in the original dataset.

\section{Related Work}
We first discuss operations on rows (to reduce the number of points $n$) and then discuss operations on columns (to reduce dimensionality of points $p$). For general surveys of big data visualization methods, see \citep{UnwinTheusHofmann,AliBigData,PenaBigData}. For more detail on feature extraction and dimensionality reduction in visual analytics, see \citep{GuyonFeatureExtraction,Fekete2002InteractiveIV,SeekAView}. 

\subsection{Reducing Rows}
First, we review row-reducing methods that use the original rows in the reduced matrix, including sampling, clustering, and squashing. Second, we introduce row-reducing methods that create new representative rows in the reduced matrix through aggregation.

\subsubsection{Sampling}
Sampling from the original data matrix is one way of reducing the sample size $n$, or the number of rows of the data matrix. A well-chosen sub-sample can represent the original dataset and recover many of the features of the original dataset.

\begin{figure}[h]
\centering
\includegraphics[width=15cm]{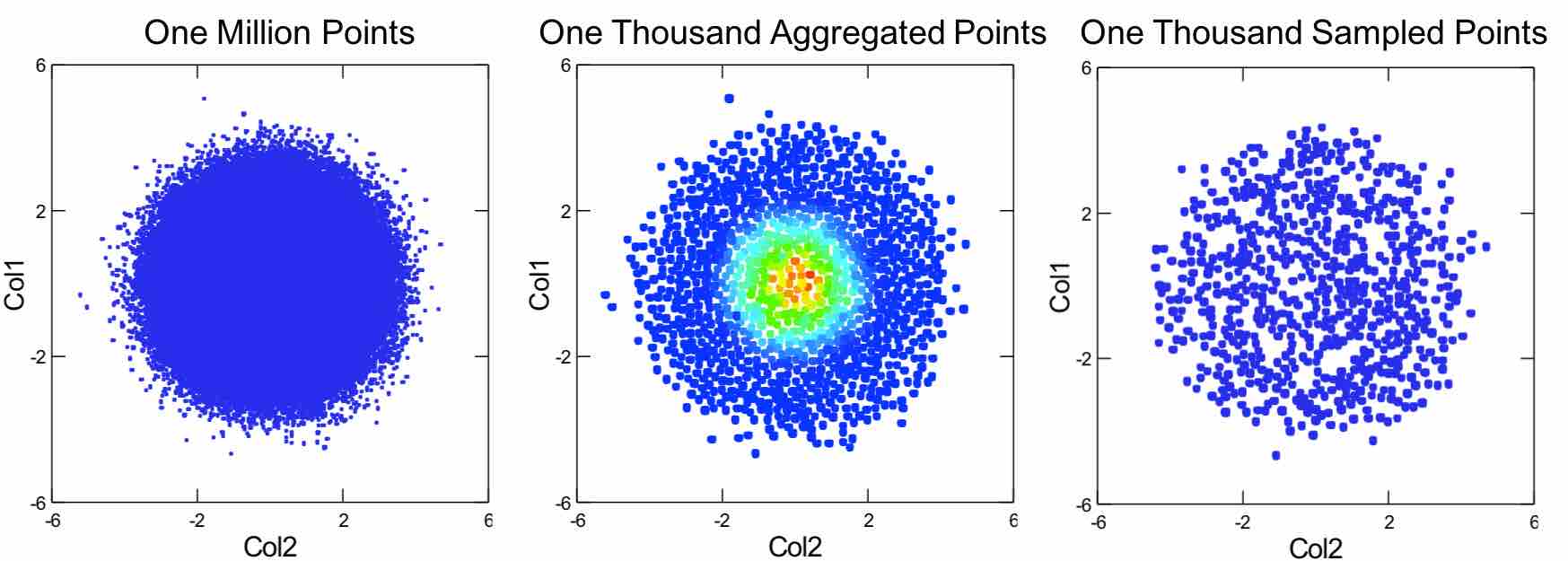}
\caption {Scatterplots of a million spherical Gaussian (Normal) random numbers (left panel), a thousand aggregated points using the Matrix Row Sketch algorithm (middle panel), and a thousand randomly sampled points (right panel). We have used color to highlight points in the middle panel that have larger frequencies resulting from aggregation. The other two plots contain single points that all have a frequency of 1. We claim that the middle panel scatterplot is a more accurate visual representation of the million-points distribution in the left panel, especially with regard to points at the periphery of the distribution. In addition, the sampled scatterplot in the right panel introduces holes that do not exist in the original or aggregated plot.}
\label{fig:summary}
\end{figure}

Sampling is not a general solution to our challenges, however. The main problem is that sampling tends to exclude statistical information in areas of low density, especially involving outliers. Unfortunately, these are the features one most wants to see when first exploring a dataset. Figure~\ref{fig:summary} illustrates this problem. In the rightmost panel of the figure, we see a simple random sample of a million point Gaussian dataset rendered in the leftmost panel. Notice that the outermost points in the raw data plot of a million points do not show up in the sampling plot. The center panel shows our sketching algorithm applied to the same dataset. All the outermost points appear in this plot. And the overall density (a bivariate spherical normal distribution) is more accurately displayed in the central panel.

\subsubsection{Clustering}

Clustering has been used for decades to reduce a large set of points to a smaller, more tractable, set suitable for subsequent analysis. In this approach, individual points are replaced by their cluster centroids and the count of original points in each cluster is recorded and used in subsequent statistical and visual analytics. 

The $k$-means algorithm is probably the most popular clustering algorithm used for this purpose. It partitions points in a space into $k$ Voronoi cells. In theory, all points in a cell are closer to the centroid of that cell than they are to any other cell centroid. Also in theory, $k$-means centroids tend to lie in a subspace specified by a principal components decomposition \citep{KMeansAndPCA,KMeansAndPCA2,ParkEtAl}.

We qualify these statements with the phrase ``in theory'' because there are many varieties of $k$-means algorithms; unfortunately, the theoretical solution is NP Hard \citep{Mahajan,Luo_etal2019}. Furthermore, identifying the dimensionality of the embedding subspace is not easy. Most $k$-means algorithms require specifying $k$ in advance of fitting. Ad hoc choices do not guarantee meaningful structure can be identified \citep{XieBigData}. And some measures of cluster-goodness appear to work well in many cases, but fail in others \citep{calinskiHarabasz,Silhouettes}

\subsubsection{Squashing}
For the purpose of retaining statistical information when $n$ is large, DuMouchel and colleagues coined the term DataSquashing to describe algorithms that attempt to preserve statistical information when ``flattening'' flat files \citep{DataSquashing}. Other similar approaches are reviewed in \citet{Squashing}. These methods work well for certain classes of distributions. They do not claim to be an overall solution to the problem when $n$ is large, however. And it is not clear how it can be applied to reduce $p$ either.

\subsubsection{Aggregation}
An alternative way of reducing the sample size is aggregating. Unlike sampling, aggregating does not necessarily select data points from the original dataset. Instead, it creates new representative data points based on the distribution, geometry, or topology of the original dataset. 

\paragraph{1-dimensional Aggregation}
The simplest, and probably oldest, form of aggregation involves a single variable. 
Histogramming is a simple method for aggregating values on a single variable.

\begin{enumerate}
\begin{singlespace}
\item Choose a small bin width ($k = 500$ bins works well for most display resolutions).
\item Bin rows in one pass through the data.
\item When finished, average the values in each bin to get a single centroid value.
\item Delete empty bins and return centroids and counts in each bin. 
\end{singlespace}
\end{enumerate}

An alternative one-dimensional aggregation method is based on dot plots \citep{dotplots}. Instead of choosing bins of equal width, we stack dots of radius $r$ to represent points. We choose $r$ to result in $k$ stacks; smaller values of $r$ yield more stacks. This algorithm is a one-dimensional version of the row sketching algorithm introduced in this paper.
\textit{Vector quantization} involves dividing a set of points into exclusive subsets. It is equivalent to histogramming with equal or unequal bin widths.

\paragraph{2-dimensional Aggregation}
Two-dimensional aggregation is a simple extension of the one-dimensional histogram algorithm. We take a pair of columns to get $(x,y)$ tuples and then bin them into a $k \times k$ rectangular grid. After binning, we delete empty bins and return centroids based on the averages of the coordinates of members in each grid cell. 

Although a little more expensive to compute, \textit{hexagonal bins} \citep{KosugiHexBin,CarrLittlefield} are preferable to rectangular binning in two dimensions. With square bins, the distance from the bin center to the farthest point on the bin edge is larger than that to the nearest point in the neighboring bin. The square bin shape leads to local anisotropy and creates visible Moir\'e patterns. Hexagonal binning reduces this effect. 

The surface of a sphere is a two-dimensional object. Consequently, we can bin $(x, y)$ tuples on a globe. It seems reasonable to select hexagons to tile the globe, but a complete tiling of the sphere with hexagons is impossible. Compromises are available, however. \citet{CarrISEA} and \citet{Kimerling} discuss this in more detail.

\paragraph{$n$-dimensional Aggregation}
Unfortunately, high-dimensional aggregation cannot be accomplished through simple extensions of 2D binning. Tiling high-dimensional spaces is problematic \citep{Sayood}. 

A number of papers attack the high-dimensional problem through 2D slices of the nD data. One of the best applications is also one of the oldest: the Grand Tour \citep{GrandTour}. A more recent implementation follows a different route through space based on Hamiltonian paths \citep{HurleyOldfordEulerianTours}. Both smoothly animate the path of 2D projections and give users the chance to control the process. But a collection of 2D binnings across an nD space does not accurately reflect or necessarily reveal joint structures in $n$ dimensions. Other subspace aggregations share similar problems. Preaggregated hash tables, for example, can improve response time from databases, but they are useful mainly for low-dimensional tables \citep{PahinsBigData}. Other approaches include machine-guided subspace views \citep{AutomatedAnalysisForViz,SeekAView,luo_li_SPCA_2021}.

\subsection{Reducing Columns}
In this section, we review column-reducing methods that output the original columns in the reduced matrix. Then we review column-reducing methods that creates new representative columns in the reduced matrix through projections.

\subsubsection{Feature Extraction}
Feature extraction involves finding a subset of features (columns) that can be used in subsequent analytics or visualizations. Most of these methods involve cluster analysis (in one form or another) or principal components \citep{GuyonFeatureExtraction,DyBrodley,FriedmanMeulman,Cheng1999EntropybasedSC}. These two classes of analytics are used to decompose variation in order to identify columns most contributing to that subspace variation. 

Unlike projection methods we will introduce below, and like matrix sketching, these approaches aim to produce sets of original variables rather than composites so that interpretation of results can be made in the original data space. However, some feature extraction methods may not depend on statistical procedures like cluster analysis or \textit{PCA} and sometimes they do not rest on assumptions concerning the distribution of the data.

\subsubsection{Projection}

A projection, in the restricted sense employed in most visual analytics, is the mapping of a set of points in $p$ dimensions to a subspace of $k$ dimensions. This kind of projection can involve a linear map (as with principal components) or a nonlinear map (as in manifold learning). These methods can also be used for modeling (as in unsupervised learning) or as preliminary steps in feature engineering \citep{Cavallo,TatuKeim}.

\paragraph{Linear projection}
Principal Component Analysis (PCA) is among the most popular linear dimension reduction methods. The original statistical algorithm for computing principal components \citep{PCA,HotellingPCA}, begins with a covariance matrix $S$ derived from the product of a column-centered $X$ matrix and its transpose and computes an eigendecomposition of that matrix (equation~\ref{eq1}), 
 
\begin{equation} \label{eq1}
\begin{split}
S & = X^TX / n = VDV^T%\\
   %& = VDV^T
\end{split}
\end{equation}
where the $D$ is a diagonal matrix of eigenvalues and $V$ is a matrix of eigenvectors. The PCA projects the data matrix linearly to the principal component directions. 

Principal components were originally designed for multivariate normally distributed points with zero mean. They can still be useful for snapshots of other types of data. More recent methods like nonlinear principal components \citep{NonlinearPCA} and sparse principal components \citep{SparsePCA} are more flexible.

The Singular Value Decomposition (SVD) works directly on a rectangular matrix. It is a generalization of the Principal Components decomposition, as equation~\ref{eq2} shows. The SVD algorithm bypasses the need for computing the covariance matrix $S$.
\begin{equation} \label{eq2}
\begin{split}
X & = UDV^T \\
 X^TX & = (UDV^T)^TUDV^T = VD^TU^TUDV^T  = VD^2V^T \\
\end{split}
\end{equation}

Linear projections may not preserve metrics, and therefore geometry of the dataset \citep{HLuoGeneralized2020}. A major group of existing visualization procedures when $p$ is large make use of dimension reduction via linear projections. However, projections often violate metric axioms -- points close together in higher-dimensional space may be far apart in lower-dimensional projections. Conversely, points far apart in higher-dimensional space may be close together in a projection. 

\paragraph{Nonlinear Projection.}
The development of nonmetric multidimensional scaling of symmetric similarity or dissimilarity matrices in the 1960's \citep{ShepardA,ShepardB,Kruskal1964} led to interest in projecting rectangular data into low-dimensional subspaces. Caroll and Chang at Bell Laboratories developed a topological embedding program called \textit{PARAMAP} \citep{PARAMAP}. Since then, numerous researchers have developed models and programs along similar lines  \citep{isomap,BelkinNiyogi,RoweisSaul,tSNE,UMAP}. 

These \textit{manifold learning} methods, linear or nonlinear, assume points lie near a $k$-dimensional manifold embedded in a subspace of the $p$-dimensional ambient space and they assume the conditional distribution of the distances of points to the manifold (residuals) is random and relatively homogeneous. Unlike principal components, however, manifold learning methods do their best when the distribution of errors is close to a manifold; they do not do well with data containing substantial error.

Nonlinear projections, although attempting to preserve local geometric features, can also violate metric axioms, which can be harmful. For example, a low-dimensional map from a higher-dimensional space that induces viewers to infer that \textit{dissimilar} points in the higher-dimensional space are \textit{similar} can lead to false conclusions. Conversely, a low-dimensional map from a higher-dimensional space that induces viewers to infer that \textit{similar} points in the higher-dimensional space are \textit{dissimilar} can lead to false conclusions. This is a commonly-expressed warning in the manifold learning community \citep{wattenberg2016how}.

\subsection{Reducing Rows and Columns}
Matrix Sketching \citep{SimpleMatrixSketching} reduces rows or columns or both simultaneously. Our algorithm is a form of matrix sketching. In general, matrix sketching represents a matrix $X$ through a sketch matrix $\tilde X$ such that the error  $\left\lVert X^TX - \tilde X ^T \tilde X \right\rVert$ is relatively small.
\subsubsection{\textit{CUR} Decompositions}
We exemplify the matrix sketching method using the CUR decomposition. The \textit{CUR} decomposition is similar to the Singular Value Decomposition (\textit{SVD}) \citep{Stewart98fouralgorithms,CURpaper}:
\begin{equation} \label{eq3}
X = CUR \\
 \end{equation}

For $X_{np}$ the \textit{CUR} components are dimensioned as $C_{nk}$, $U_{km}$ and $R_{mp}$, where $1 \le m \le n$ and $1 \le k \le p$. This restriction forces equation \ref{eq3} to be an approximation rather than an isometry. Some \textit{CUR} algorithms are relatively efficient \citep{CURpaper}; their time performance can be $O(np^2)$. 

The same kind of rank reduction can be achieved with an \textit{SVD}. Unlike \textit{SVD}s, however, the matrices $C$ and $R$ are subsets of the rows and columns of $X$ rather than linear combinations of them. This feature of \textit{CUR}, shared by our matrix sketching algorithm, is distinctive.  

However, \textit{CUR} differs from our algorithm in at least one important respect. \textit{CUR} outputs three matrices while ours outputs one. For a more general survey of matrix decompositions suited to high-dimensional data, see \citep{probabilisticMatrixDecompositions}. 

\section{A New Matrix Sketching Algorithm}
Our matrix sketch is based on two associated algorithms. We first consider sketching rows to reduce $n$. Then we consider sketching columns to reduce $p$.
\subsection{Sketching Rows}
Our algorithm for reducing the number of rows is related to a greedy approximate solution to the \textit{k-center problem} \citep{k-center}:
\begin{quote}
Given $n$ points in $p$-dimensional metric space and an integer $k \le n$, find the minimum radius $r$ and a set of balls of radius $r$ centered on each of $m$ points such that all $n$ points lie within the union of these balls.
\end{quote}

Instead of conditioning on $k$, however, we condition on $r$ and denote the final number of centers to be $m$ in following discussion. Our algorithm is derived from a variant of the Leader clustering algorithm, which is described in \citet{ClusteringAlgorithms}. Algorithm~\ref{RowSketcher Algorithm} shows the rows sketching algorithm.

\begin{algorithm}
\caption{RowSketcher Algorithm}
\label{RowSketcher Algorithm}
\KwData{$X_{n,p}$ (data matrix)}
\KwIn{$r = .25 / (\log{n})^{1/p}$ (default value)}
\SetKwFunction{EuclideanDistance}{EuclideanDistance}
\EuclideanDistance ($a,b$) is a Euclidean distance function\\
\KwResult{$exemplars$ (list of ball centers), $members$ (list of lists)}
\Begin{
     $m = 1$ \\
     $row$ (array of length $p$) = first row of $X$\\
     $exemplars$ = new list, initialized to contain $row$\\
     $members$ = new list of lists, initialized to contain empty list\\
}
\For{$i = 1, \dots, n$} {
     $newExemplar = true$\\
     $row = X[i,.]$ ($i$th row of $X$)\\
      \For{$j = 1, \dots, m$} {
           $d = \EuclideanDistance(row, exemplars[j])$\\
           \If {$d < r$} {
               add $i$ to $members[j]$\\
               $newExemplar = false$\\
               $break$\\
            } 
       }
       \If{newExemplar} {
             $m = m + 1$\\
             add $row$ to $exemplars[m]$\\
             add new list to $members[m]$, initialized with $i$
        }
}
\end{algorithm}

\pagebreak
\subsubsection{Notes on RowSketcher Algorithm}

\begin{enumerate}[noitemsep]
\item The default value of $r$ is designed roughly to be below the expected value of the distances between $n(n-1)/2$ pairs of points distributed randomly in a $p$ dimensional unit hypercube. Increase $r$ to produce fewer exemplars, decrease to produce more. 
\item The $exemplars$ list contains a list of row values representing points defining exemplar neighborhoods.
\item The $members$ list of lists contains one list of indices for each exemplar pointing to members of that exemplar's neighborhood.
\item A consequence of aggregating rows is that statistics on the aggregates must include frequency weighting. This requirement rules out some statistical libraries that do not incorporate frequencies. Most statistical packages for survey analysis, for example, incorporate frequency weighting in their basic statistics (or, in the case of SAS, SPSS, SYSTAT, or STATA, everywhere). 
\item The Leader algorithm \citep{ClusteringAlgorithms} creates exemplar-neighborhoods in one pass through the data. It is equivalent to centering balls in $p$ dimensional space on points in the dataset that are considered to be exemplars. Unlike $k$-means clustering, the Leader algorithm 
\begin{enumerate}[noitemsep,nolistsep]
\item centers balls on actual data points rather than on centroids of clusters,
\item constrains every ball to the same radius rather than allowing clusters to have different diameters,
\item involves only one pass through the data rather than iterating to convergence via multiple passes,
\item produces many balls rather than a few clusters. 
\end{enumerate}

\item In rare instances, the resulting exemplars and members can be dependent on the order of the data, but not enough to affect the description of the joint density of points because of the large number of exemplars produced. We are characterizing a high-dimensional density by covering it with many small balls. Even relatively tight clusters produced by a clustering algorithm will be chopped into pieces by the Leader algorithm. Nevertheless, if this is a concern, we can visit the list of exemplars in random order on each iteration.

\item The time complexity of Algorithm~\ref{RowSketcher Algorithm} is $O(nmp)$.
\end{enumerate}

\begin{figure}[h!]
\centering{\includegraphics[width=250pt]{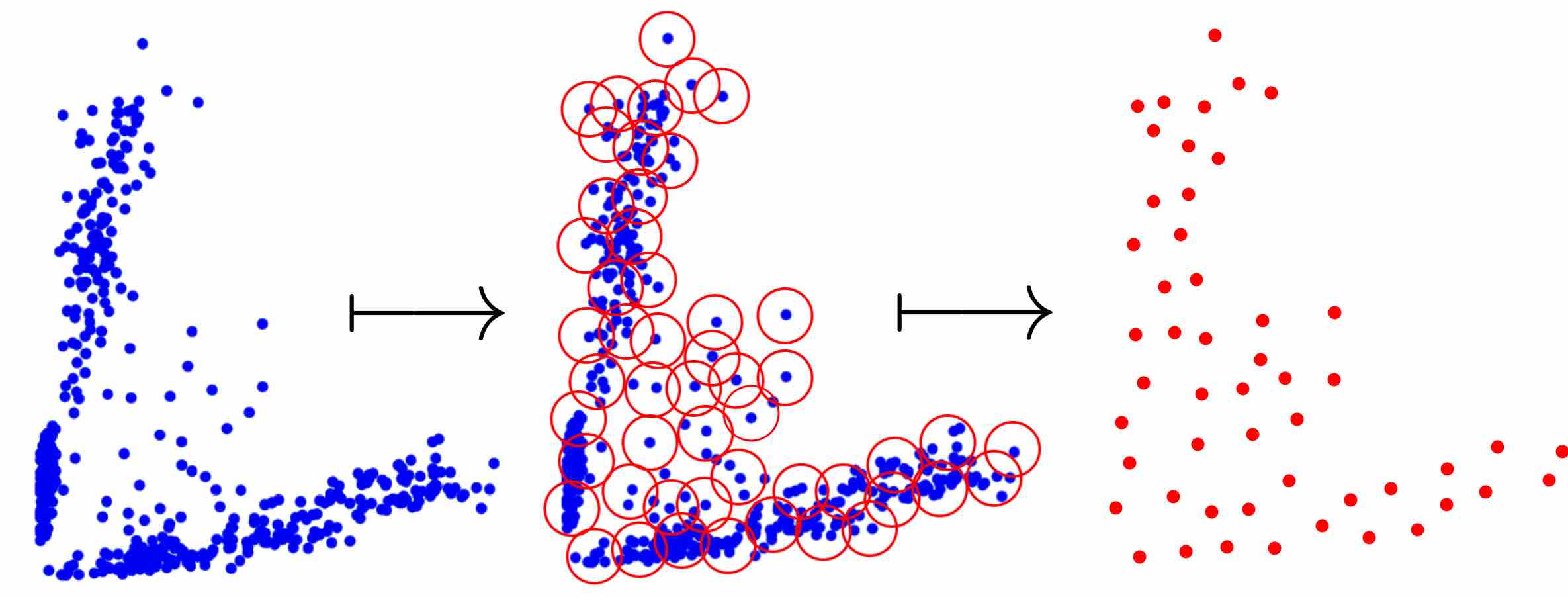}}
\caption{\label{fig:2D} Visualization of the row sketching algorithm in a 2D space.}
\end{figure}

Figure ~\ref{fig:2D} shows a schematic depicting a 2D implementation of our algorithm.

\subsubsection{Categorical Variables}
To incorporate categorical variables in the RowSketcher algorithm, we need to convert categories to numerical values. Correspondence Analysis (CA) \citep{CA,MCA} suits our purpose. We begin by representing a categorical variable with a set of dummy codes, one code (1 or 0) for each category. These codes comprise a matrix of 1's and 0's with as many columns as there are categories for that variable. We then compute a principal components decomposition of the covariance matrix of the dummy codes. This analysis is done separately for each of $k$ categorical variables in a dataset. CA scores on the rows are computed for each categorical variable by multiplying the dummy codes on that row's variable times the eigenvectors of the decomposition for that variable. Computing the decomposition separately for each categorical variable is equivalent to doing a multiple correspondence analysis separately for each variable instead of pooling all the categorical variable dummy codes into one matrix. This application of CA to deal with the visualization of nominal data was first presented in \citet{RosarioWard}. 

Unfortunately, this approach loses the mapping of rows to specific categories. When members are assigned to exemplars, the category values of the exemplars must be chosen. If we are interested in grouping by category, we must implement a different approach. Namely, we maintain categories in separate cells (e.g., hash table entries) and then apply RowSketcher separately to the continuous values in each cell. If one or more categorical variables have high cardinality, however, this approach becomes increasingly impractical.

\subsubsection{Performance of RowSketcher Algorithm}
Table~\ref{millions} is based on the million Gaussians dataset used for Figure~\ref{fig:summary}. The three stub panels show the results of computing basic statistics on three different datasets. The Sketch Dataset uses our RowSketcher algorithm (with radius = .119) and the Random Sample uses a simple random sample to extract 200 rows from the million. Both row-reducing methods do quite well overall, but there is a  striking difference. RowSketcher yields the maxima and minima of the raw dataset, but the Sample method does not. In short, RowSketcher, coupled with frequency-weighted formulas, does quite well on basic statistics.

\begin{table}[h!]
\begin{center}
\begin{tabular}{rrrr}
& \textbf{X} & \textbf{Y} & \textbf{Z} \\
\hline
\textbf{Original Dataset}\\
m & 1,000,000 & 1,000,000 & 1,000,000\\
n & 1,000,000 & 1,000,000 & 1,000,000\\
Min  & -4.683 & -5.208  & -5.184\\
Max  & 5.061 & 4.702  & 4.832 \\
Mean & -0.001 & 0.000 & 0.001 \\
Median & 0.000 & 0.001 & 0.001 \\
SD & 1.000 & 1.000 & 1.001 \\
\hline
\textbf{Sketch Dataset}\\
m & 200 & 200 & 200\\
n & 1,000,000 & 1,000,000 & 1,000,000 \\
Min  & -4.683 & -5.208  & -5.184\\
Max  & 5.061 & 4.702  & 4.832 \\
Mean & 0.009 & 0.001 & -0.003\\
Median & -0.007 & 0.076 & 0.031 \\
SD & 1.104 & 1.124 & 1.127 \\
\hline
\textbf{Random Sample}\\
m & 200 & 200 & 200\\
n & 1,000,000 & 1,000,000 & 1,000,000 \\
Min  & -2.916 & -2.539  & -2.628\\
Max  & 2.950 & 2.513  & 3.002 \\
Mean & -0.049 & -0.020 & -0.015 \\
Median & -0.048 & 0.013 & 0.010 \\
SD & 1.011 & 0.939 & 1.046 \\
\hline
\end{tabular}
\end{center}
\caption{\label{millions}Original vs. Row Sketch and Random Sample of a Million Gaussians.}
\end{table}

To test additional capabilities of our algorithms, we generated and tested six datasets against perhaps the most widely used row-and-column sketching algorithm -- the CUR decomposition. We did not cherry pick these datasets; they were the only ones we generated in order to test capabilities we considered important for visualization sketches. 

The first two rows of Table~\ref{comparisons} show the results of comparisons between the Row Sketch algorithm and the CUR algorithm. For these comparisons, we coded the Linear Time CUR algorithm \citep{LinearCUR} in Java and ran the Java versions of our sketching algorithms. 

For the Outlier2D test, we generated a thousand bivariate normal points in two dimensions ($\rho = .8, \sigma = .1$) and added an outlier at (.6, .6). We then asked both algorithms to sketch the dataset down to 500 rows. The RowSketcher algorithm included the outlier, but the CUR algorithm did not.

For the Inlier2D test, we generated a thousand points aligned on a unit circle with a polar conditional Normal standard deviation of .1. The resulting scatterplot resembled a donut sprinkled with random Normal points on its surface. In addition, we added one more point at (0, 0) in the center of the donut that we called an ``inlier.'' As before, we asked both algorithms to sketch the dataset down to 500 rows. The RowSketcher algorithm included the inlier, but the CUR algorithm did not. 

For exploratory visualization, any sketching algorithm should capture anomalous points (outliers and inliers) so that we can analyze them before doing statistical modeling. Because of the distance properties inherent in our Leader algorithm, this is a likely result and desirable property of our row sketching.

\begin{table}[h!]
\begin{center}
\begin{tabular}{rrrrrrrr}
& \textbf{Rows}  & \textbf{Cols} & \textbf{rows}  & \textbf{cols} & \textbf{CPU} & \textbf{Corr} & \textbf{Hit}\\
\hline
\textbf{Outlier2D Dataset}\\
RowSketcher &1000 & 2 & 500 & 2 & 39ms & NA & Yes\\
CUR &1000 & 2 & 500 & 2  & 53ms & NA & No\\
\hline
\textbf{Inlier2D Dataset}\\
RowSketcher &1000 & 2 & 500 & 2 & 36ms & NA & Yes\\
CUR &1000 & 2 & 500 & 2  & 59ms & NA & No\\
\hline
\textbf{Cluster Dataset}\\
ColSketcher &1000 & 100 & 1000 & 2 & 887ms & .99 & Yes\\
CUR &1000 & 100 & 1000 & 2  & 355ms & .82 & No\\
\hline
\textbf{Donut Dataset}\\
ColSketcher &1000 & 100 & 1000 & 2 & 886ms & .96 & Yes\\
CUR &1000 & 100 & 1000 & 2  & 486ms & .68 & No\\
\hline
\textbf{Outlier Dataset}\\
ColSketcher &1000 & 100 & 1000 & 2 & 899ms & .98 & Yes\\
CUR &1000 & 100 & 1000 & 2  & 360ms & .53 & No\\
\hline
\textbf{SwissRoll Dataset}\\
ColSketcher &1000 & 100 & 1000 & 3 & 1091ms & .94 & Yes\\
CUR &1000 & 100 & 1000 & 3  & 593ms & .53 & No\\
\hline
\end{tabular}
\end{center}
\caption{\label{comparisons}Performance of Sketcher vs. CUR algorithm. Capital Rows and Cols contain the original dimensions of the tables generated. Lower-case rows and cols contain the dimensions of the sketched dataset. CPU shows the time to run the algorithms on a 2.8 GHz Intel Core i7 Macbook Pro running macOS High Sierra. Corr shows the Frobenius correlation of the sketch (relevant only for column sketching). Hit shows whether the respective algorithms recovered the critical rows or columns being tested.}
\end{table}

\subsection{Sketching Columns}
Algorithm~\ref{ColSketcher Algorithm} contains the ColSketcher algorithm. 

\begin{algorithm}
\caption{ColSketcher Algorithm}
\label{ColSketcher Algorithm}
\KwData{$X_{m,p}$ (data matrix sketched with algorithm~\ref{RowSketcher Algorithm})}
\KwIn{$maxCorrelation = .95$ (default value)}
\SetKwFunction{summ}{sum}
\SetKwFunction{FrobeniusCorrelation}{FrobeniusCorrelation}
\SetKwFunction{dist}{dist}
\summ ($a,b$) is an element-wise sum function on two arrays\\
\dist ($X_{m,p}$) produces an $m(m-1)/2$ array of squared distances\\
\FrobeniusCorrelation $(a,b) = (a \boldsymbol{\cdot} b) / (\|a\|\|b\|)$ \\
\KwResult{$selectedCols$ (list of selected columns)}
%\begin{algorithmic}[1]
\Begin {
    $correlation = 0.0$\\
     $selectedCols = $ list of selected columns initialized empty\\
     $previousColDist = m(m-1)/2$ length array of zeros\\
     $allColDist = m(m-1)/2$ array of distances from \dist($X_{.,.}$)\\ 
}
\While{$correlation < maxCorrelation$} {
     $bestColumn = 0$ \\
     $bestCorrelation = 0.0$ \\
     $previousBestCorrelation = 0.0$ \\
     \For{$j = 1, \dots, p$} {
          \If{$j \notin selectedCols$} {
                $jColDist = \dist(X[.,j])$\\
                $cumColDist = \summ(jColDist, previousColDist)$\\
                $correlation = \FrobeniusCorrelation(cumColDist, allColDist)$\\
                \If{$correlation > bestCorrelation$} {
                     $bestColumn = j$\\
                     $bestCorrelation = correlation$\\
                }
             }
          }
         $bestColDist = \dist(X[., bestColumn])$\\
         $previousColDist = \summ(bestColDist, previousColDist)$\\
         add $bestColumn$ to $selectedCols$\\
         $previousBestCorrelation = bestCorrelation$\\
}
\end{algorithm}

\subsubsection{Notes on ColSketcher Algorithm}
\begin{enumerate}[noitemsep]
\item Our column sketching algorithm is based on a variant of greedy forward feature selection with a Frobenius coefficient to measure similarity between distance matrices.
\item Our algorithm accumulates squared distances in $cumColDist$ on each iteration over $p$. This saves time by confining distance computations at each step to a pair of column arrays rather than to a matrix of columns. Since squared distances are additive, we need to compute only one-dimensional distances in each step and cumulate them with previously-computed squared distances.
\item The time complexity of Algorithm~\ref{ColSketcher Algorithm} is $O(kmp)$.
\end{enumerate}

\subsubsection{Performance of ColSketcher Algorithm}
We present in this section various approaches to evaluating the performance of the column sketching algorithm. Some of these are designed to illustrate differences from other algorithms rather than overall effectiveness.

\paragraph{Feature Detection.}
The last four rows of Table~\ref{comparisons} show the results of comparisons between our Column Sketch algorithm and the CUR algorithm. For the Cluster test, we generated a thousand N(0, .1) points in 100 dimensions. In two of these dimensions (columns) we randomly centered each of these points on one of six cluster centroids located on a $3 \times 3$ grid. We then asked each algorithm to reduce the 100 columns to 2 columns. Our Column Sketcher algorithm correctly located these two columns. The CUR algorithm did not.

For the Donut Dataset test, we generated the same donut we used for the row sketching test, but this time we embedded it in only two of the 100 dimensions. As before, we asked each algorithm to reduce the 100 columns to 2 columns. Our Column Sketcher algorithm correctly located these two embedding columns. The CUR algorithm did not.

For the Outlier dataset, we generated a thousand bivariate Normal points ($\rho = .8, \sigma = 1$), but we embedded them in only two of the 100 dimensions. The remaining columns contained coordinates based on random Gaussians having standard deviations of .1. We then added one outlier in the bivariate target columns pair at (6, 6). We then asked each algorithm to reduce the 100 columns to 2 columns. Our Column Sketcher algorithm correctly located these two columns. The CUR algorithm did not.

For the SwissRoll dataset, we generated a thousand points lying on a Swiss Roll manifold embedded in 3 dimensions \citep{RoweisSaul}. The remaining 97 dimensions consisted of standard Normal N(0, 1) variates. The test on this dataset was to see if a sketch algorithm could identify the subspace embedding the roll without being led astray by the random error. Our Column Sketcher algorithm correctly located these three columns. The CUR algorithm did not.

\paragraph{Distance Preservation.}
The loss functions and error bounds for many of these projection methods are variously based on the discrepancy between the coordinates of the points in the low-dimensional embedding space and their coordinates in the higher-dimensional residual space. The loss function for \textit{ColSketcher}, by contrast, is based on the correlation between the distances reproduced by the sketch matrix and the original distances between points. 
In short, most popular projection methods, with the exception of multidimensional scaling, which is challenging to apply on large datasets \citep{Paradis}, are not distance-preserving. For details, \citet{Compadre} have developed a tool to examine distance preservation visually.

\begin{figure}[h]
\centering{\includegraphics[width=400pt]{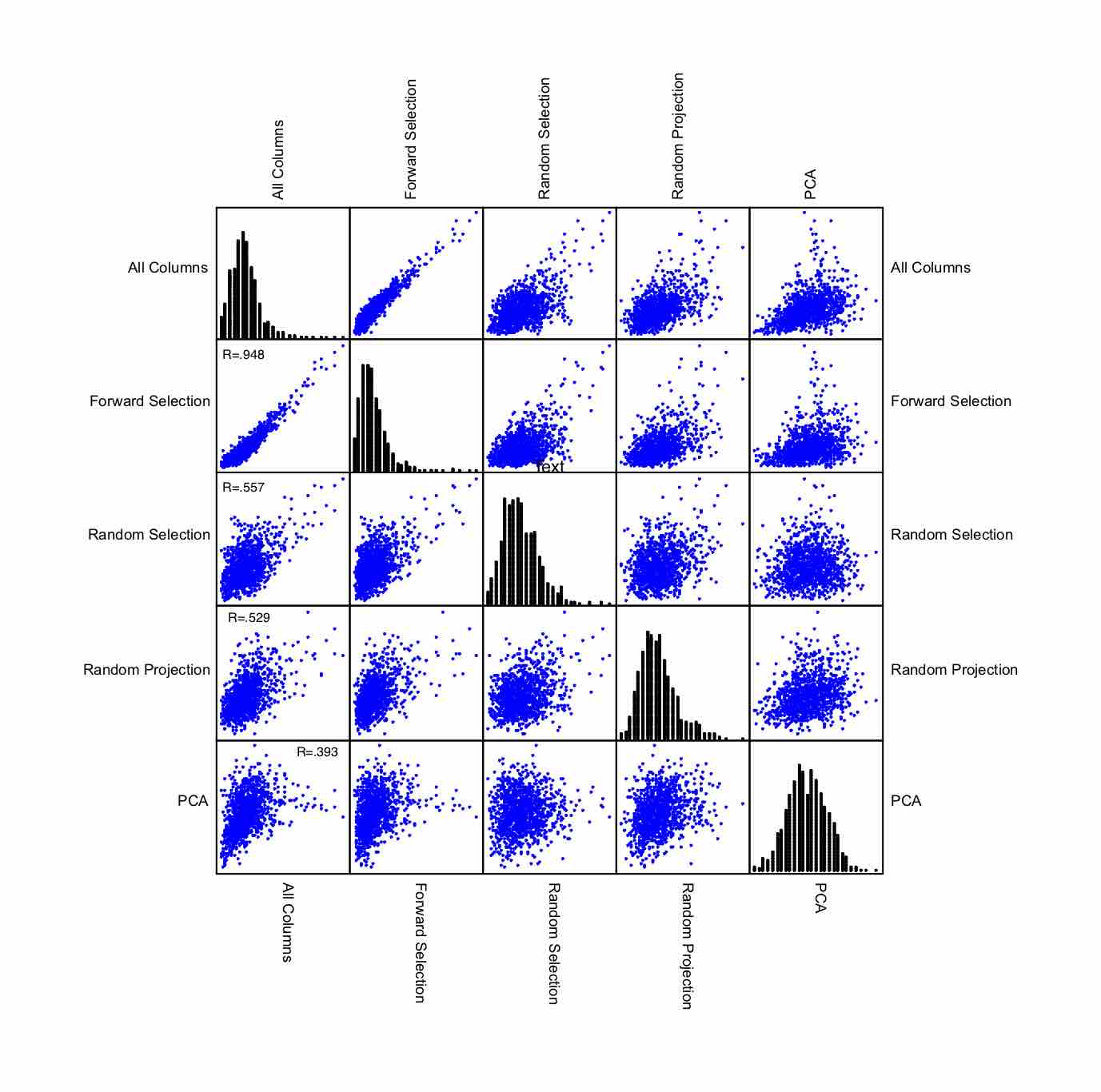}}
\caption{\label{fig:distances} Comparison of column reducing algorithms using the gene expression dataset. The points in the plots represent all pairwise distances between points in higher-dimensional space (20,531 dimensions) and pairwise distances between the same points in projected space (40 dimensions). The forward selection ColSketcher is the only algorithm in this collection that substantially preserves relative distances among points.}
\end{figure}

We can compare the distance-preserving capabilities of several column-reducing algorithms.
Figure ~\ref{fig:distances} shows a SPLOM of the column sketch algorithm vs. several other projection methods. The data are taken from the gene expression dataset used in Figure~\ref{fig:heatmap}. In all methods, we reduced 20,531 columns to 40 columns. While this reduction might not have been optimal for all methods, it allows us to compare the preservation of distances after the same amount of reduction. The results are dramatic. Clearly, the column sketch outperforms the other methods. Incidentally, we omitted manifold learning methods because they are not distance preserving algorithms; they are designed to upweight short distances and downweight long ones.

\begin{figure}[h]
\centering{\includegraphics[width=300pt]{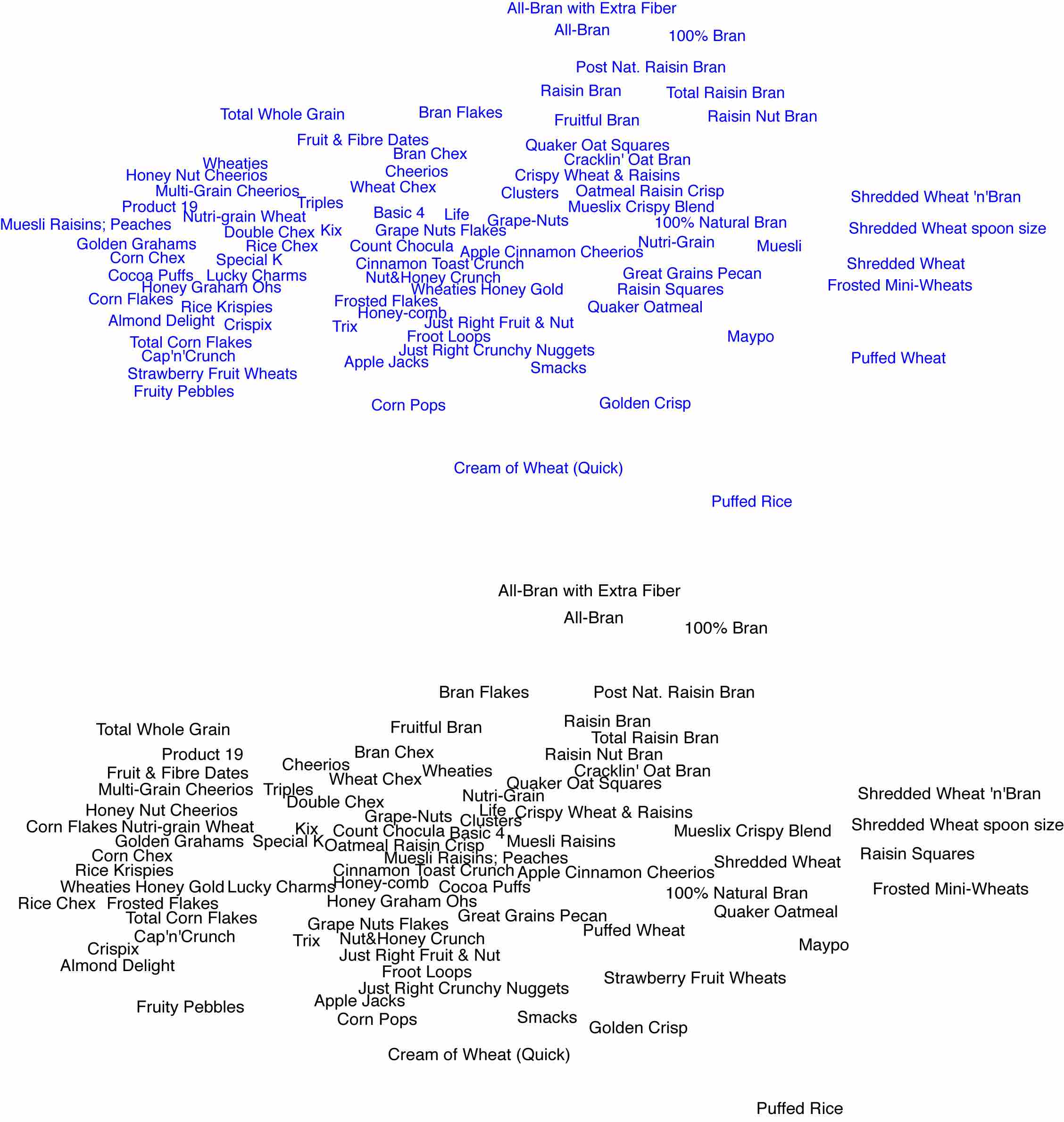}}
\caption{\label{fig:cereals} Comparison of analyses based on the column sketch algorithm (lower panel) and all columns (upper panel) using a cereals dataset from kaggle.}
\end{figure}

\paragraph{Accuracy of ColSketcher}
A simple test of the column sketching algorithm is to compute two analyses of the same data -- one on the full dataset and the other on the column-sketched dataset. Figure~\ref{fig:cereals} shows two multidimensional scalings. The upper panel shows the scaling of 77 cereals from a kaggle dataset (\url{https://www.kaggle.com/crawford/80-cereals}). There are 13 continuous variables represented by the columns (calories, protein, fat, sodium, fiber, carbo, sugars, potassium, vitamins, shelf, weight, cups, rating). We computed Euclidean distances among the cereals based on all 13 columns and then did an \textit{MDS} on the resulting distance matrix. The cereals are colored blue in this coordinate plot.

The lower panel, in black, shows the scaling of the same cereals using seven columns selected by the column sketch algorithm (calories, sodium, fiber, sugars, potassium, vitamins, shelf). The Frobenius correlation between the row distances in the full dataset and in the sketch dataset is 0.98. While there are some differences in detail, the result of the sketch algorithm is visibly close to the result based on all the variables. This will not be necessarily true if we use analytic visualization methods that do not preserve distances (such as principal components, \textit{tSNE}, or \textit{UMAP}).
 
\begin{figure*}[h]
\centering{\includegraphics[width=400pt]{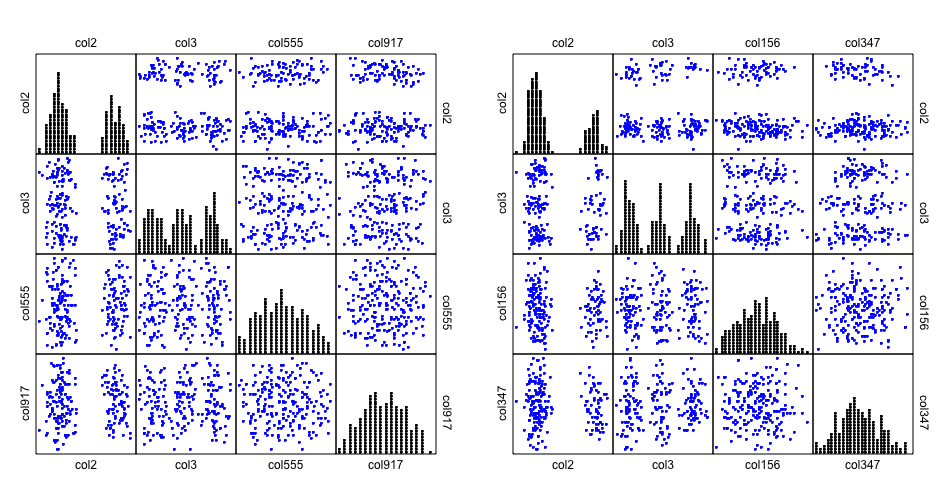}}
\caption{\label{fig:gaussians} SPLOM on 1000  rows and 1000 Gaussian columns reduced to 4. }
\end{figure*}

\paragraph{Sketching Approximately Square matrices}
Figure~\ref{fig:gaussians} shows scatterplot matrices on sketched rows and columns of an artificial dataset. We generated 1,000 independent Gaussians on each of 1,000 variables. For the third and fourth variables, we generated two and three Gaussians respectively, separated into clusters. For the left panel, we ran RowSketcher and then ran ColSketcher on the output from RowSketcher. For the right panel, we ran the two sketchers in the opposite order. In both cases, we forced it to select four variables out of the 1,000. Figure~\ref{fig:gaussians} shows that either sketcher orderings selected the two anomalous variables, col2 and col3. The additional scatterplots show the remaining patterns that are embedded in this multivariate dataset. Any of the additional variables would have revealed the same patterns when plotted against each other or against col2 and col3.

\subsection{Visualization}
This section presents several multivariate visualizations that are particularly suited to our matrix sketching algorithm on rows and columns of the data matrix. 

\subsubsection{Biplots}

\begin{figure}[h]
\centering{\includegraphics[width=400pt]{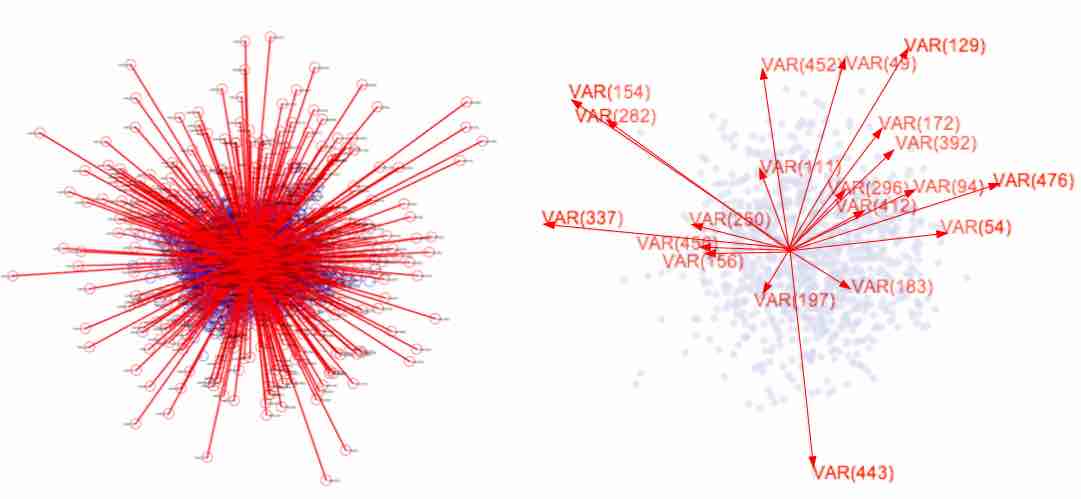}}
\caption{\label{fig:biplot} Full Madelon dataset biplot (left) and matrix sketched dataset biplot (right).}
\end{figure}

Figure~\ref{fig:biplot} shows a biplot~\citep{biplot} of the Madelon training dataset \citep{Madelon}. We reduced 2000 rows to 1000 and 500 columns to 20. This biplot represents principal component loadings with vectors (red) and scores with points (blue) -- all in the same frame. The biplot in the left panel obscures most of the variation in cases and variables. The biplot in the right panel shows the 20 vectors representing the column variables. The canonical correlation between the coordinates of the vectors in the right plot with the coordinates of the corresponding vectors in the left plot is 0.89. This indicates that the right plot is accurately representing the relevant loadings in the principal components of the full dataset. In addition, the sketch plot spans the full 2D space the way the full-data plot does. It is not a seriously biased representation.

\subsubsection{Parallel Coordinates}
Parallel Coordinates, in various forms, are one of the most popular multidimensional visual analytic methods for big data \citep{Inselberg,Zhang20162DAM,ClusteredPC}. Their well-known weakness is visual clutter, caused both by many cases and many variables. Obvious remedies for this problem include the use of kernels, profile aggregation, and sorting of variables to reduce crossings. Our sketching algorithm on rows and columns makes all these remedies easier to realize, especially with limited computational resources.

\begin{figure}[h]
\centering{\includegraphics[width=450pt]{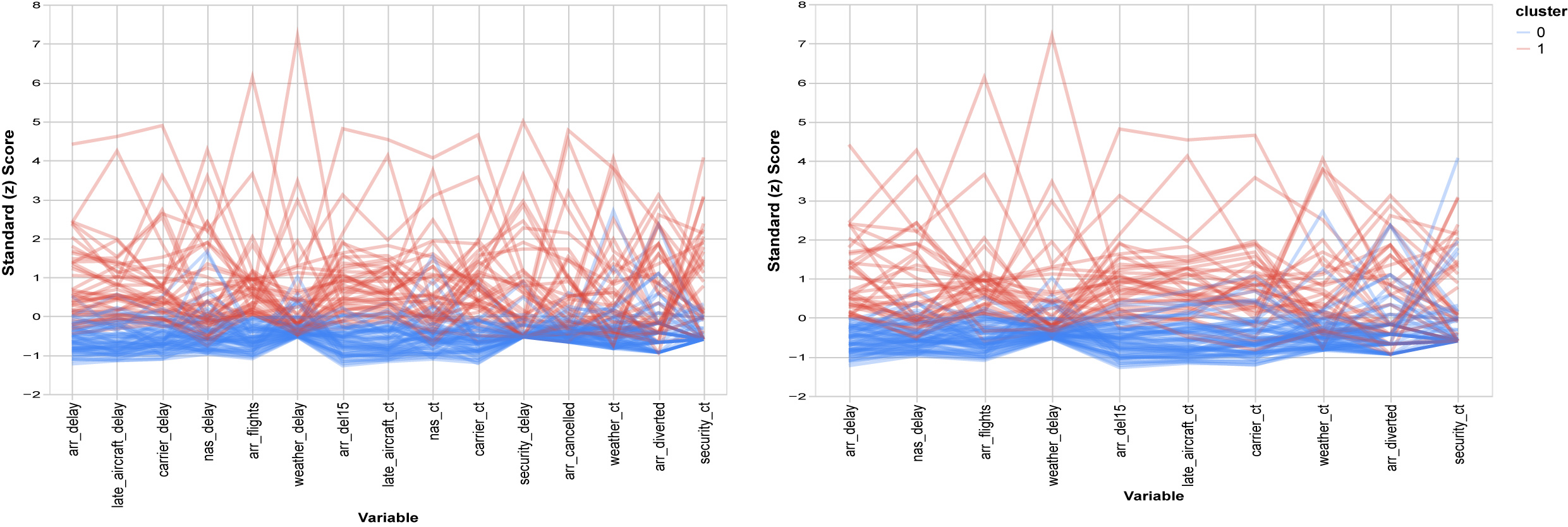}}
\caption{\label{fig:airlineDelays} Parallel Coordinates plot of 1762 rows reduced to 100 and 15 columns reduced to 10 columns}
\end{figure}

Figure~\ref{fig:airlineDelays} shows parallel coordinates using our sketching algorithm on rows and columns on a popular dataset comprising delays in air traffic performance (\url{https://www.transtats.bts.gov/OT_Delay/OT_DelayCause1.asp}). A $k$-means cluster analysis is used to color the display, which clearly reveals two different clusters \citep{calinskiHarabasz} on the performance variables. The panel on the left is the plot before sketching. The RowSketcher and ColumnSketcher algorithms applied in the right panel reduce the clutter and, importantly, preserve the overall structure.

\subsubsection{Heatmaps}
Cluster heatmaps involve joint reorderings of the rows and columns of a data matrix using hierarchical clustering \citep{wilkinsonFriendly}. They are impractical for big data for two reasons. First, display resolution prevents the rendering of cells in large heatmaps, even when cells are depicted in single pixels. Second, the number of rows and/or columns in big data matrices exceeds the computational efficiency of hierarchical clustering. Matrix sketching is suited as a remedy for these problems.

\begin{figure}[h]
\centering{\includegraphics[width=250pt]{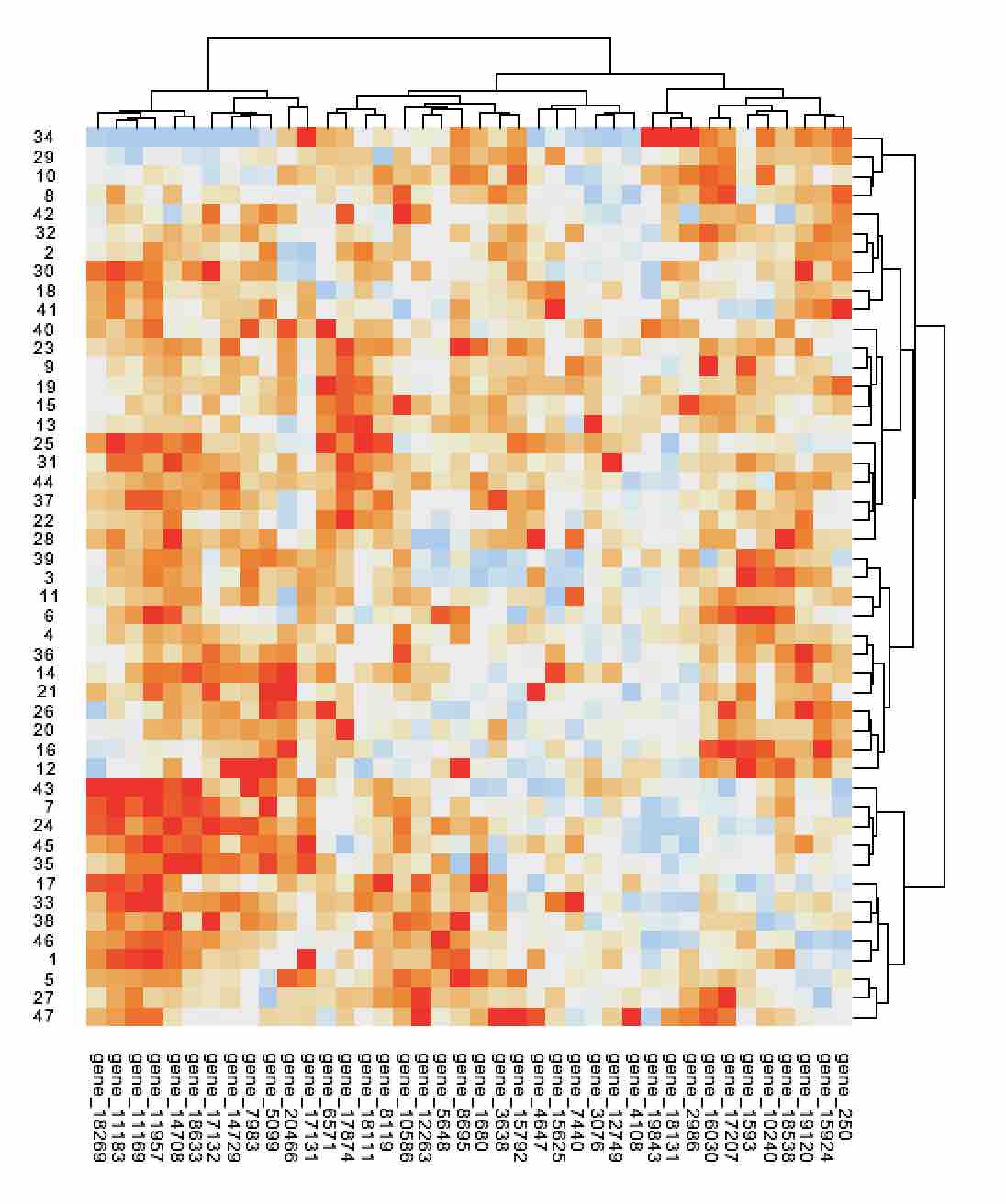}}
\caption{\label{fig:heatmap} Heatmap of matrix-sketched gene expression data.}
\end{figure}

Figure~\ref{fig:heatmap} shows a cluster heatmap of gene expression data using matrix sketching. The data are from \citet{WeinsteinHeatmap} %\url{https://archive.ics.uci.edu/ml/datasets/gene+expression+cancer+RNA-Seq}
, see also \citet{Khomtchouk}. %\url{https://journals.plos.org/plosone/article?id=10.1371/journal.pone.0176334}
We reduced 801 rows and 20,531 columns to 47 rows and 40 columns. The display indicates joint clusters (particularly in the lower left) that might be fruitful for further analysis. 

\pagebreak
\subsubsection{Scatterplot Matrices}
\begin{figure*}[h]
\centering{\includegraphics[width=500pt]{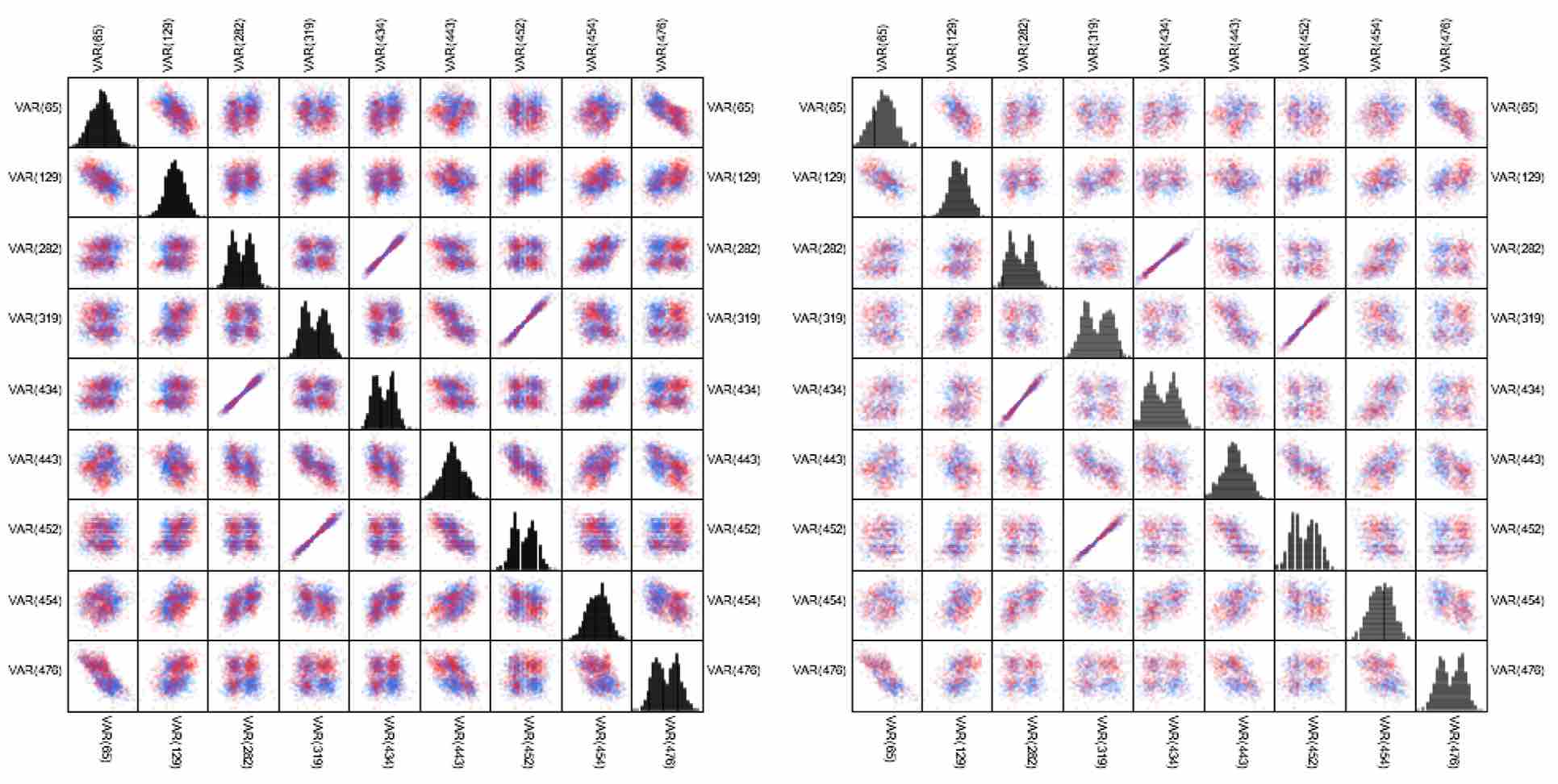}}
\caption{\label{fig:sploms} SPLOM on 2000 Madelon rows (left) and 1000 (right), 500 columns reduced to 10. The tenth variable (CLASS) is used to color the scatterplots.}
\end{figure*}

Figure~\ref{fig:sploms} shows a scatterplot matrix on sketched columns of the Madelon dataset \citep{Madelon}. There are two interesting aspects of these plots. First, the sketched columns reveal anomalous artificial structures embedded in this dataset. In particular, the two straight-line relationships between columns clearly stand out against the other patterns. These are the only anomalous ones of this kind in the whole dataset. Sketching is not an anomaly detector, but when embedded among relatively homogeneous distributions, anomalous relations are likely to be exposed. Columns on which there are outliers, for example, will have more leverage in the distance correlation calculations. Second, the two SPLOMs are remarkably similar. 

\subsubsection{Boxplots}

\begin{figure}[h]
\centering{\includegraphics[width=400pt]{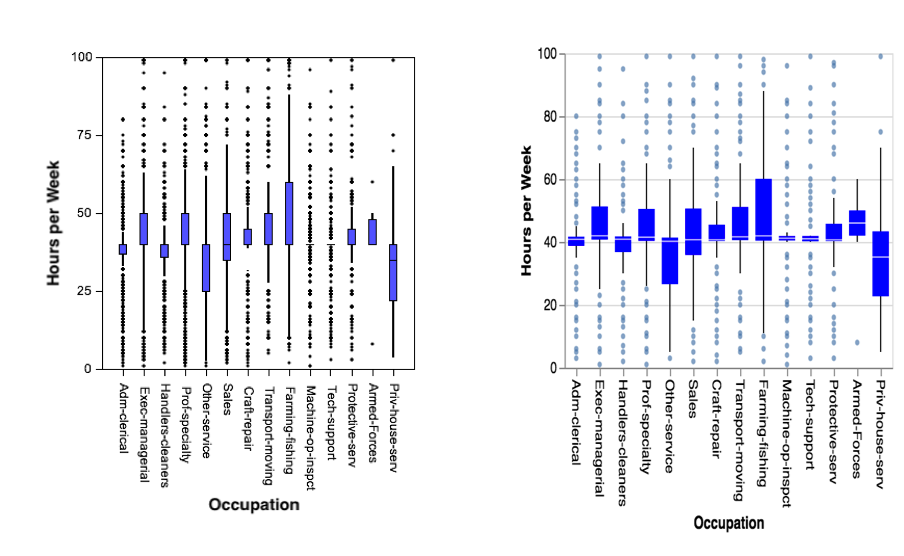}}
\caption{\label{fig:boxplot} Full Adult dataset boxplot (left, $N$ = 32,000) and row sketched Adult dataset boxplot (right, $n$ = 430).}
\end{figure}

Figure~\ref{fig:boxplot} shows a Tukey schematic (boxplot) for the U.S. Census Adult dataset (\url{https://archive.ics.uci.edu/ml/datasets/adult}). The plot on the left is for the complete dataset and the one on the right is derived from the row sketch of the same dataset. Despite a nearly 75 percent reduction in the number of cases, the two plots are visually almost indistinguishable. Furthermore, the row sketched boxplot outliers are brushable as long as the members indices are retained as pointers.

\subsubsection{Nonlinear Manifolds}

\begin{figure}[h]
\centering{\includegraphics[width=400pt]{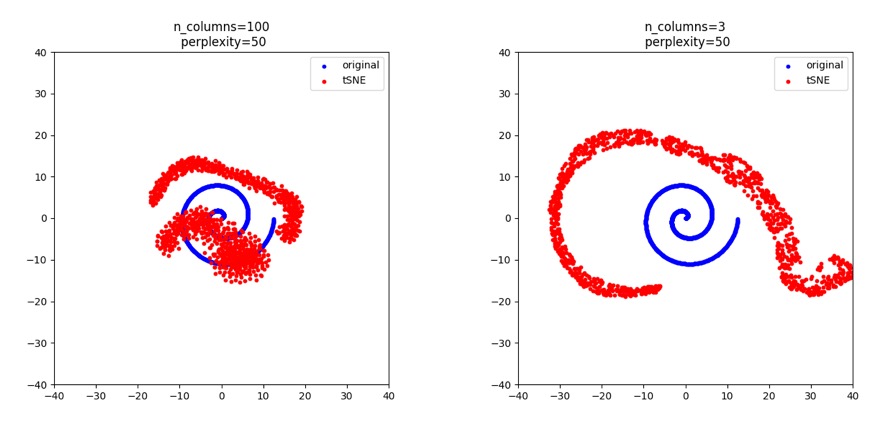}}
\caption{\label{fig:manifold} t-{SNE} scalings of a thousand-points Swiss Roll dataset.  The left panel shows a t-{SNE} projection based on all 100 variates; the Swiss Roll manifold has been incorrectly split in two. The right panel shows a t-{SNE} projection based on the first three variates selected by our ColSketcher algorithm.}
\end{figure}

A use for ColSketcher that we did not originally anticipate was as a preprocessor for manifold learning algorithms like t-SNE. Manifold learning algorithms are well-known to lack robustness against large amounts of error in ambient dimensions.  The SwissRoll dataset test in Table~\ref{comparisons} demonstrated that ColSketcher might be useful for locating embedding dimensions in wide datasets containing substantial error. Indeed, Figure~\ref{fig:manifold} shows how this capability can be leveraged to assist algorithms like t-SNE in handling these difficult problems. The plot on the left shows the result of a t-SNE projection of the entire dataset. It incorrectly breaks the manifold into two parts. The plot on the right, by contrast, shows a t-SNE projection based on the column-sketched version of the dataset. The Swiss Roll is correctly rendered as a single manifold. An additional benefit of using our sketchers is to make practical the application of iterative manifold methods and other computationally expensive algorithms to larger datasets.

\pagebreak
\section{Conclusion}
In a landmark paper relatively unknown to many computer scientists and statisticians today, Amos Tversky discussed the use of real vector spaces in data science \citep{TverskyFeaturesOfSimilarity}. At the time of the paper, psychologists were enthusiastic about the possibility of using multidimensional scaling (\textit{MDS}) to derive a cognitive map (points in a metric space) from judgments of the similarities between objects. Tversky demonstrated that some types of data are inappropriate for methods that depend on metric axioms. A simple example is the triad of statements most observers would agree with:
\begin{itemize}[noitemsep]
\itemsep0em 
\item Miami is similar to Havana
\item Havana is similar to Moscow
\item But Miami is not similar to Moscow
\end{itemize}

Tversky argued against the indiscriminate use of metric space models in psychology, but there is perhaps a wider range of indiscriminate usage of nonlinear manifold models in machine learning and visualization today. Users of these methods may assume that they are appropriate for any numerical data. 

A corollary of Tversky's observation, in the context of today's multidimensional visualization practices, might be our point in the Related Work section that visualizations that violate metric axioms can be harmful, leading viewers to misinterpret similarities between objects. Judging dissimilar points as similar or similar points as dissimilar can lead to false conclusions. Today's popular multidimensional visualization algorithms are not intrinsically flawed; their flaws lie in their indiscriminate uses that do not take into account the assumptions underlying them. 

In addition to the primary motivation for this research (distance-preservation under projections), there is a significant concomitant benefit. It involves reification of composites. As \citet{CURpaper} point out, 
\begin{quotation}
Although the truncated \textit{SVD} is widely used, the vectors $u^i$ and $v^i$ themselves may lack any meaning in terms of the field from which the data are drawn. For example, the eigenvector\\
\indent
 $[(1/2)age - (1/\sqrt{2})height + (1/2)income]$,\\
being one of the significant uncorrelated ``factors'' or ``features'' from a dataset of people's features, is not particularly informative or meaningful. This fact should not be surprising. After all, the singular vectors are mathematical abstractions that can be calculated for any data matrix. They are not ``things'' with a ``physical'' reality.

Nevertheless, data analysts often fall prey to a temptation for reification, i.e., for assigning a physical meaning or interpretation to all large singular components. 
\end{quotation}

\citet{CURpaper} explain axis-parallel representation as a strength of the \textit{CUR} decomposition. We agree. While our algorithm is fundamentally different from theirs, it shares interpretive advantages with \textit{CUR}.

We do not propose that our sketching algorithm on rows and columns replace other methods for handling big data problems. Each method has its own advantages. Our algorithm on rows and columns has several, each designed to facilitate visual analysis of large datasets. First, it returns a subset of a given matrix, not a set of additive composites. This facilitates brushing and linking to real data values rather than to composites. Second, our algorithm is more scalable than other projection methods, especially iterative ones like manifold learning or projection pursuit. And, finally, our algorithm is distance-preserving so that the resulting low-dimensional visualizations are less likely to violate the metric axioms when we use sketching inside the visualization flow running from data to perceived structures, patterns, and relationships.
\subsection*{Acknowledgment}
Supporting materials, including source code, are available at \newline \url{https://github.com/hrluo/DistancePreservingMatrixSketch}. Wilkinson devised the row and column algorithms, wrote the main section of the paper, coded the Java applications and the dataset evaluations. Luo devised the proofs, wrote the Appendix, coded the R and Python versions, and edited the paper. The authors especially thank one of the reviewers for valuable suggestions.
\pagebreak
\begin{center}
{\large\bf SUPPLEMENTAL MATERIALS}
\end{center}

\appendix
\section{\label{ErrBound}Error Bounds and Distance Preserving Properties}

\subsection{Row Algorithm Error Bound}

For the row (Leader) algorithm, the error for exemplar-to-exemplar
distances is zero because the exemplars are original data points. 

For member-to-member, the worst case error for a single pair is 2$r$,
where each point is at the furthest boundary of the ball in which
it is a member relative to the other point in the pair, which is at
the furthest point in its own ball. Since the estimate of the distance
between the two points is based on the respective exemplars, 2$r$
is the worst possible error in this case. 

\begin{figure*}[h]
\centering{\includegraphics[width=300pt]{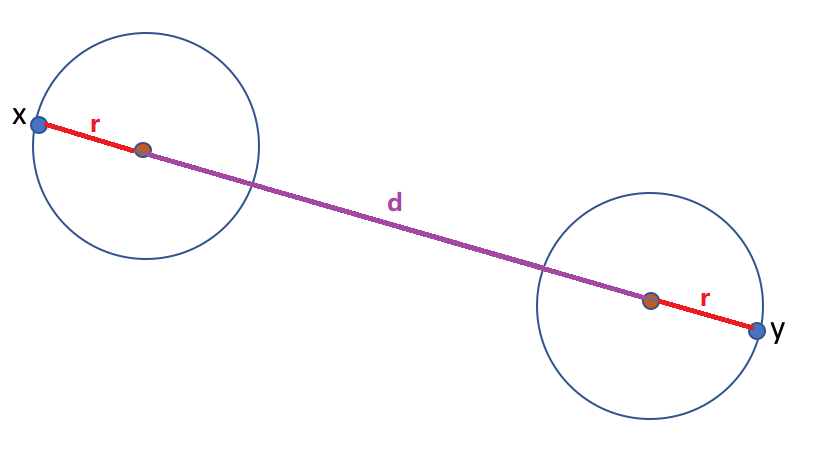}}
\caption{\label{fig:rowDistances} Distances between exemplars and extremes in 2 dimension.}
\end{figure*}

For example, in Figure~\ref{fig:rowDistances}, the true distance between points
$x$ and $y$ is $r+d+r$. However, then in the reduced dataset, distance
between points $x$ and $y$ will be the inter exemplar distance $d$.
Therefore, the error between the pairwise distance of $x,y$ in the original
and reduced dataset is 2$r$. 

Pairwise distances are the only concern in the row sketching algorithm.
Since we fix the radius $r$ of each exemplar, we need to know the
minimal number of exemplars we needed for an $r$-covering number for the
dataset represented by the $n\times p$ data matrix $\boldsymbol{X}$.
If we know the distribution of the dataset, we could integrate
this over all points to get the worst case overall error and expected
error. 

If the variances of the columns are significantly different, we can normalize each column of the $n\times p$
data matrix $\boldsymbol{X}$ such that each entry in $\boldsymbol{X}$
is in $[0,1]$. In this case, each squared pairwise distance $d\left(\boldsymbol{X}(i,:),\boldsymbol{X}(j,:)\right)^{2}$
is in $[0,p]$. Instead of removing rows from $\boldsymbol{X}$ in
the row algorithm, we simply replace the rows of data points with the
row of corresponding exemplar to obtain reduced $n\times p$ matrix
$\boldsymbol{X}_{(R)}$. Only the nonrepetitive rows represent data
points.

\begin{align*}
d\left(\boldsymbol{X}(i,:),\boldsymbol{X}(j,:)\right)^{2}-d\left(\boldsymbol{X}_{(R)}(i,:),\boldsymbol{X}_{(R)}(j,:)\right)^{2} & =\left[d\left(\boldsymbol{X}(i,:),\boldsymbol{X}(j,:)\right)-d\left(\boldsymbol{X}_{(R)}(i,:),\boldsymbol{X}_{(R)}(j,:)\right)\right]\cdot\\
 & \left[d\left(\boldsymbol{X}(i,:),\boldsymbol{X}(j,:)\right)+d\left(\boldsymbol{X}_{(R)}(i,:),\boldsymbol{X}_{(R)}(j,:)\right)\right]\\
 & \leq2r\cdot2\sqrt{p}\\
 & \leq4\sqrt{p}r
\end{align*}

\begin{thm}
Consider the normalized data matrix $\boldsymbol{X}$. We use the
notation $\boldsymbol{X}_{ij}$ to denote the $(i,j)$-th entry; $\boldsymbol{X}(i,:)$
to denote the $i$-th row;$\boldsymbol{X}(:,j)$ to denote the $j$-th
column of the $m\times p$ data matrix $\boldsymbol{X}$, and we use
subscript $\boldsymbol{X}_{(R)}$ to denote the distance matrix after
row reduction. Then
\[
\max_{i,j}\left|d\left(\boldsymbol{X}(i,:),\boldsymbol{X}(j,:)\right)^{2}-d\left(\boldsymbol{X}_{(R)}(i,:),\boldsymbol{X}_{(R)}(j,:)\right)^{2}\right|\leq4\sqrt{p}r,
\]
\end{thm}

Note that the distances between exemplars are unchanged and the differences
between these pairs are zero. Suppose there are $K(r)\leq n$ exemplars,
then there are $(n-K(r))\cdot(n-K(r)-1)$ pairs of non-exemplars,
and $K(r)(n-K(r))$ exemplar-non-exemplar pairs. The overall error
can be described by 
\begin{align*}
& \sum_{i,j}\left|d\left(\boldsymbol{X}(i,:),\boldsymbol{X}(j,:)\right)^{2}-d\left(\boldsymbol{X}_{(R)}(i,:),\boldsymbol{X}_{(R)}(j,:)\right)^{2}\right| \\
& \leq4\sqrt{p}r\cdot\left[(n-K(r))\cdot(n-K(r)-1)+K(r)(n-K(r))\right].\\
 & =4\sqrt{p}r\cdot(n-K(r))\cdot(n-1)
\end{align*}

Although $K(r)$ is non-increasing in $r$ (i.e. $n-K(r)$ is non-decreasing),
we do not have an explicit expression of the quantity $K(r)$. However,
we can sometimes have a probabilistic estimate when the distributional
properties of $\boldsymbol{X}$ are known. 

Empirically, for a fixed $n$, if $r$ is small enough, then the row algorithm produces
negligible error for all points and the overall error depends mostly
on the column algorithm. For which we will discuss in length below. 

\subsection{Column Algorithm Error Bound}

\subsubsection{Column Sketching with Frobenius coefficient}

When we use the Frobenius coefficient to measure the similarity between
squared distance matrices, an error bound for the column algorithm
involving $\epsilon$ could be derived as below.

The Frobenius matrix inner product between two $m\times p$ matrices
$\boldsymbol{A},\boldsymbol{B}$ is defined as $\left\langle \boldsymbol{A},\boldsymbol{B}\right\rangle \coloneqq\text{tr}\left(\boldsymbol{A}^{T}\boldsymbol{B}\right)$
for the space consisting of all $m\times p$ matrices is an inner
space, denoted as $\mathbb{R}^{m\times p}$. With this notion of inner
product and its induced Frobenius norm $\|\boldsymbol{A}\|_{F}\coloneqq\text{tr}\left(\boldsymbol{A}^{T}\boldsymbol{A}\right)$,
we can define the cosine between two matrices $\boldsymbol{A},\boldsymbol{B}\in\mathbb{R}^{m\times p}$
as $\cos\left(\boldsymbol{A},\boldsymbol{B}\right)\coloneqq\frac{\left\langle \boldsymbol{A},\boldsymbol{B}\right\rangle }{\sqrt{\left\langle \boldsymbol{A},\boldsymbol{A}\right\rangle \cdot\left\langle \boldsymbol{B},\boldsymbol{B}\right\rangle }}$.

We use the notation $\boldsymbol{X}_{ij}$ to denote the $(i,j)$-th
entry; $\boldsymbol{X}(i,:)$ to denote the $i$-th row;$\boldsymbol{X}(:,j)$
to denote the $j$-th column of the $m\times p$ data matrix $\boldsymbol{X}$,
and we use subscript $\boldsymbol{D}_{(k)}$ to denote the distance
matrix in the $k$-th loop.

The column algorithm starts with a \emph{squared distance matrix}
$\boldsymbol{D}_{(0)}=\boldsymbol{0}\in\mathbb{R}^{m^{2}\times m^{2}}$
(we assume a full distance matrix instead of an $\frac{m(m-1)}{2}$
array) with all entries being zero. We can compute $\boldsymbol{D}(j)=\mathtt{dist}\left(\boldsymbol{X}(:,j)\right),j=1,\cdots,p$
to be squared distance matrices for the $j$-th column of the data
matrix $\boldsymbol{X}$. This indexing shall not be confused with
our subscript convention above.

Then, we choose one column from the data matrix $\boldsymbol{X}$
such that the $\max_{j=1,\dots,p}\cos\left(\boldsymbol{D_{X}},\boldsymbol{D}(j)\right)$
is attained by using the squared distance matrix of this column. In
the space $\mathbb{R}^{m^{2}\times m^{2}}$, this choice is equivalent
to choosing from ``vectors'' $\boldsymbol{D}(1),\cdots,\boldsymbol{D}(p)$
such that the cosine is maximized. Geometrically, we choose one vector
in the space $\mathbb{R}^{m^{2}\times m^{2}}$ which has the smallest
angle against $\boldsymbol{D_{X}}$. Then we update the $\boldsymbol{D}_{(0)}$
to be $\boldsymbol{D}_{(1)}$. By the Pythagorean theorem, the $(i,j)$-th
entry of $\boldsymbol{D_{X}}$ is $\|\boldsymbol{X}(i,:)-\boldsymbol{X}(j,:)\|_{2}^{2}$,
which can be decomposed into $\sum_{k=1,\cdots,p}|\boldsymbol{X}_{ik}-\boldsymbol{X}_{jk}|^{2}=\sum_{k=1,\cdots,p}\boldsymbol{D}(k)_{ij}$
due to orthogonality. Examine each entry in both distance matrices,
we have 
\[
\boldsymbol{D_{X}}=\sum_{k=1,\cdots,p}\boldsymbol{D}(k),
\]
which explains why the $\boldsymbol{D}_{(0)},\boldsymbol{D}_{(1)},\cdots$
are updated additively in each loop. 

In short, we start from a zero distance matrix $\boldsymbol{D}_{(0)}$
and subsequently choose from $\boldsymbol{D}(1),\cdots,\boldsymbol{D}(p)$
to approach the direction of $\boldsymbol{D_{X}}$ until either we
reach $\boldsymbol{D_{X}}=\sum_{k=1,\cdots,p}\boldsymbol{D}(k)$ or
the $\boldsymbol{D}_{(C)}$ has a sufficiently small angle (measured
by threshold $\epsilon$) to $\boldsymbol{D_{X}}$.
\begin{example}
(Co-planar example) Consider a concrete example of $3\times3$ matrix
$\boldsymbol{X}=\left(\begin{array}{ccc}
0 & 1 & 2\\
0 & 4 & 5\\
0 & 6 & 9
\end{array}\right)$, which represents 3 points $(0,1,2)^{T},(0,4,5)^{T},(0,6,9)^{T}$
in $\mathbb{R}^{3}$.

In the column algorithm we have initialized $\boldsymbol{D}_{(0)}=\left(\begin{array}{ccc}
0 & 0 & 0\\
0 & 0 & 0\\
0 & 0 & 0
\end{array}\right)$, and the distance matrix for the full dataset is $\boldsymbol{D_{X}}=\mathtt{dist}(\boldsymbol{X})=\left(\begin{array}{ccc}
0 & 18 & 50\\
18 & 0 & 8\\
50 & 8 & 0
\end{array}\right)$. 

For each of the 3 columns we compute 
\begin{align*}
\boldsymbol{D}(1) & =\mathtt{dist}(\boldsymbol{X}(:,1))=\left(\begin{array}{ccc}
0 & 0 & 0\\
0 & 0 & 0\\
0 & 0 & 0
\end{array}\right), & \cos\left(\boldsymbol{D_{X}},\boldsymbol{D}(1)\right) & =0,
\end{align*}

\begin{align*}
\boldsymbol{D}(2) & =\mathtt{dist}(\boldsymbol{X}(:,2))=\left(\begin{array}{ccc}
0 & 9 & 25\\
9 & 0 & 4\\
25 & 4 & 0
\end{array}\right), & \cos\left(\boldsymbol{D_{X}},\boldsymbol{D}(2)\right) & =\frac{2888}{\sqrt{5776\times1444}}=1,
\end{align*}

\begin{align*}
\boldsymbol{D}(3) & =\mathtt{dist}(\boldsymbol{X}(:,3))=\left(\begin{array}{ccc}
0 & 9 & 49\\
9 & 0 & 16\\
49 & 16 & 0
\end{array}\right), & \cos\left(\boldsymbol{D_{X}},\boldsymbol{D}(3)\right) & =\frac{5480}{\sqrt{5776\times5476}}\approx0.97,
\end{align*}
(also note that $\boldsymbol{D_{X}}=2\cdot\boldsymbol{D}(2)$). 

In the first loop, since the distance matrix $\boldsymbol{D}(2)$
has the largest cosine (i.e., smallest angle) to $\boldsymbol{D_{X}}$,
we select the second column in the first loop and update 
\[
\boldsymbol{D}_{(1)}=\boldsymbol{D}_{(0)}+\boldsymbol{D}(2)=\left(\begin{array}{ccc}
0 & 9 & 25\\
9 & 0 & 4\\
25 & 4 & 0
\end{array}\right).
\]

Before the second loop, we can compute that $\cos\left(\boldsymbol{D_{X}},\boldsymbol{D}_{(1)}\right)=\frac{8664}{\sqrt{5776\times12996}}=1>0.95$
and this stops the column algorithm. 

The column reduced matrix will be $\boldsymbol{X}_{\text{col reduced}}=\boldsymbol{X}{}_{(1)}=\left(\begin{array}{c}
1\\
4\\
6
\end{array}\right)$ with the only retained column being the second column in the original
matrix. 

The distance matrix of the reduced matrix is $\left(\begin{array}{ccc}
0 & 9 & 25\\
9 & 0 & 4\\
25 & 4 & 0
\end{array}\right)$. 
\end{example}

This dataset is special in the sense that these 3 points are in fact
living in y-z plane. Intuitively, we can drop the first column (x-axis
in $\mathbb{R}^{3}$) without disturbing pairwise distances. However,
we also drop the third column, because all we care in the column algorithm
with Frobenius coefficient is the direction of $\boldsymbol{D_{X}}$,
not even its modulus.

\subsubsection{Error Bound}

Generally speaking, suppose we have chosen an $m\times p$ data matrix
$\boldsymbol{X}$ and a threshold $\epsilon\in(0,1]$, and the column
algorithm stops after $N\leq p$ steps.

\textbf{Step 1. Geometric distance between $\boldsymbol{D_{X}},\boldsymbol{D}_{(C)}$.}
We have stated in the sub-section and example above that the Frobenius
coefficient or the cosine has a natural geometric interpretation.
On one hand, we can assert that $\cos\left(\boldsymbol{D_{X}},\boldsymbol{D}_{(C)}\right)\geq\epsilon$
and subsequently $\sin\left(\boldsymbol{D_{X}},\boldsymbol{D}_{(C)}\right)\leq\sqrt{1-\epsilon^{2}}$
since $\sin^{2}\theta+\cos^{2}\theta=1$. The difference between $\boldsymbol{D_{X}},\boldsymbol{D}_{(C)}$
can be written as $\|\boldsymbol{D_{X}}-\boldsymbol{D}_{(C)}\|_{F}$,
by trigonometric geometry, we know that 
\[
\|\boldsymbol{D_{X}}-\boldsymbol{D}_{(C)}\|_{F}\leq\max\left(\|\boldsymbol{D_{X}}\|_{F},\|\boldsymbol{D}_{(C)}\|_{F}\right)\cdot\sin\left(\boldsymbol{D_{X}},\boldsymbol{D}_{(C)}\right)\leq\max\left(\|\boldsymbol{D_{X}}\|_{F},\|\boldsymbol{D}_{(C)}\|_{F}\right)\cdot\sqrt{1-\epsilon^{2}}.
\]
On the other hand, we know that $\|\boldsymbol{A}\|_{F}=\sqrt{\sum_{i=1}^{m}\sum_{j=1}^{n}|a_{ij}|^{2}}\geq\max_{i,j}|a_{ij}|\eqqcolon\|\boldsymbol{A}\|_{\max}$
for a matrix $\boldsymbol{A}=\left\llbracket a_{ij}\right\rrbracket $
(See, e.g., \citep{horn2012matrix}). Therefore,
\[
\|\boldsymbol{D_{X}}-\boldsymbol{D}_{(C)}\|_{\max}\leq\max\left(\|\boldsymbol{D_{X}}\|_{F},\|\boldsymbol{D}_{(C)}\|_{F}\right)\cdot\sqrt{1-\epsilon^{2}}.
\]

\textbf{Step 2. $\boldsymbol{D_{X}},\boldsymbol{D}_{(C)}$ as (normalized)
squared distance matrices.} But the $(i,j)$-th entry in $\boldsymbol{D_{X}}$
is the pairwise squared distance $d\left(\boldsymbol{X}(i,:),\boldsymbol{X}(j,:)\right)^{2}$
with Euclidean distance $d$, and the $(i,j)$-th entry in $\boldsymbol{D}_{(C)}$
is the pairwise distance $d\left(\boldsymbol{X}_{(C)}(i,:),\boldsymbol{X}_{(C)}(j,:)\right)$,
we know that 
\[
\max_{i,j}\left|d\left(\boldsymbol{X}(i,:),\boldsymbol{X}(j,:)\right)^{2}-d\left(\boldsymbol{X}_{(C)}(i,:),\boldsymbol{X}_{(C)}(j,:)\right)^{2}\right|\leq\max\left(\|\boldsymbol{D_{X}}\|_{F},\|\boldsymbol{D}_{(C)}\|_{F}\right)\cdot\sqrt{1-\epsilon^{2}},
\]
The left hand side of this inequality is the maximal difference between
pairwise squared distances calculated in the original and column reduced
dataset. 

However, our columns should have roughly equal variances or we have normalized each column of $\boldsymbol{X}$
such that each entry in $\boldsymbol{X}$ is in $[0,1]$. Note that
$\boldsymbol{D}_{(C)}$ is a submatrix of $\boldsymbol{D_{X}}$ and
each entry of $\boldsymbol{D_{X}},\boldsymbol{D}_{(C)}$ are inside
$[0,p]$. By definition, ${\displaystyle \|\boldsymbol{A}\|_{F}=\text{tr}\left(\boldsymbol{A}^{T}\boldsymbol{A}\right)=\sqrt{\sum_{i=1}^{m}\sum_{j=1}^{n}|a_{ij}|^{2}}}$
we have 
\begin{align}
    \max\left(\|\boldsymbol{D_{X}}\|_{F},\|\boldsymbol{D}_{(C)}\|_{F}\right)\leq\sqrt{\sum_{i=1}^{m}\sum_{j=1}^{m}p^{2}}=mp,
\end{align}
since each entry of both distance matrices are bounded in a unit hyper-cube with dimension at most $p$.
\begin{thm}
\label{thm:Given-an-}Given an $m\times p$ data matrix $\boldsymbol{X}$,
we supply it as the input of column algorithm with threshold $\epsilon\in(0,1]$.
We denote the output matrix of column algorithm as $\boldsymbol{X}_{(C)}$,
where $0\leq N\leq p$ is the number of loops until stopping. The
notation $\boldsymbol{D_{X}}=\left\llbracket d\left(\boldsymbol{X}(i,:),\boldsymbol{X}(j,:)\right)\right\rrbracket $
and $\boldsymbol{D}_{(C)}=\left\llbracket d\left(\boldsymbol{X}_{(C)}(i,:),\boldsymbol{X}_{(C)}(j,:)\right)\right\rrbracket $.
Then, 
\[
\max_{i,j}\left|d\left(\boldsymbol{X}(i,:),\boldsymbol{X}(j,:)\right)^{2}-d\left(\boldsymbol{X}_{(C)}(i,:),\boldsymbol{X}_{(C)}(j,:)\right)^{2}\right|\leq C_1\cdot\sqrt{1-\epsilon^{2}},
\]
where $C_1$ is a positive constant s.t. $C_1\geq mp$. 
\end{thm}

Since both $\boldsymbol{D_{X}},\boldsymbol{D}_{(C)}$ are symmetric
interval matrices \citep{Hladik} with entries in $[0,p]$
with column normalization, a sharper constant $C$ could be arithmetically
obtained. Let us try this error bound with the co-planar example again.
\begin{example}
(Normalized co-planar example) Consider again the example of $3\times3$
matrix $\boldsymbol{X}=\left(\begin{array}{ccc}
0 & 1 & 2\\
0 & 4 & 5\\
0 & 6 & 9
\end{array}\right)$, which represents 3 points $(0,1,2)^{T},(0,4,5)^{T},(0,6,9)^{T}$
in $\mathbb{R}^{3}$. Let's normalize columns of this data matrix
into $\tilde{\boldsymbol{X}}=\left(\begin{array}{ccc}
0 & 1/11 & 2/16\\
0 & 4/11 & 5/16\\
0 & 6/11 & 9/16
\end{array}\right)$

In the column algorithm we have initialized $\boldsymbol{D}_{(0)}=\left(\begin{array}{ccc}
0 & 0 & 0\\
0 & 0 & 0\\
0 & 0 & 0
\end{array}\right)$, and the distance matrix for the full dataset is $\boldsymbol{D_{X}}=\mathtt{dist}(\boldsymbol{X})=\left(\begin{array}{ccc}
0 & 0.11 & 0.40\\
0.11 & 0 & 0.10\\
0.40 & 0.10 & 0
\end{array}\right)$ (rounded-off to the second decimal). 

For each of the 3 columns we compute 
\begin{align*}
\boldsymbol{D}(1) & =\mathtt{dist}(\boldsymbol{X}(:,1))=\left(\begin{array}{ccc}
0 & 0 & 0\\
0 & 0 & 0\\
0 & 0 & 0
\end{array}\right), & \cos\left(\boldsymbol{D_{X}},\boldsymbol{D}(1)\right) & =0.00,
\end{align*}

\begin{align*}
\boldsymbol{D}(2) & =\mathtt{dist}(\boldsymbol{X}(:,2))=\left(\begin{array}{ccc}
0 & 0.07 & 0.21\\
0.07 & 0 & 0.03\\
0.21 & 0.03 & 0
\end{array}\right), & \cos\left(\boldsymbol{D_{X}},\boldsymbol{D}(2)\right) & \approx0.994101,
\end{align*}

\begin{align*}
\boldsymbol{D}(3) & =\mathtt{dist}(\boldsymbol{X}(:,3))=\left(\begin{array}{ccc}
0 & 0.04 & 0.19\\
0.04 & 0 & 0.06\\
0.19 & 0.06 & 0
\end{array}\right), & \cos\left(\boldsymbol{D_{X}},\boldsymbol{D}(3)\right) & \approx0.9930334,
\end{align*}

In the first loop, since the distance matrix $\boldsymbol{D}(2)$
has the largest cosine (i.e., smallest angle) to $\boldsymbol{D_{X}}$,
we select the second column in the first loop and update 
\[
\boldsymbol{D}_{(1)}=\boldsymbol{D}_{(0)}+\boldsymbol{D}(2)=\left(\begin{array}{ccc}
0 & 0.07 & 0.21\\
0.07 & 0 & 0.03\\
0.21 & 0.03 & 0
\end{array}\right).
\]

Before the second loop, we can compute that $\cos\left(\boldsymbol{D_{X}},\boldsymbol{D}_{(1)}\right)=1.00>0.95$
and this stops the column algorithm with only $\boldsymbol{X}(:,1)$
retained. 

The error bound obtain with $C=m\sqrt{p}$ is $m\sqrt{p}\cdot\sqrt{1-\epsilon^{2}}=3\cdot3\cdot\sqrt{1-0.95^{2}}\approx2.810249$, and the difference
\begin{align}
\max_{i,j}\left|d\left(\boldsymbol{X}(i,:),\boldsymbol{X}(j,:)\right)^{2}-d\left(\boldsymbol{X}_{(1)}(i,:),\boldsymbol{X}_{(1)}(j,:)\right)^{2}\right|\approx0.19140625
\end{align}
is attained for $i=3,j=1$, which is clearly bounded by $2.810249$.
\end{example}
The error bound also serves as a theoretical support for the claimed
``distance-preserving property'' of column sketching, since the
difference between $\boldsymbol{D_{X}},\boldsymbol{D}_{(C)}$ is essentially
bounded by a quantity only depending on $m,p$ and threshold $\epsilon$. 

Interestingly, in this example, even if we choose $\epsilon=0.99$,
the result still holds with one loop, but with a tighter bound
$0.733$. However, if we set the $\epsilon=0.999$ then the error bound
becomes $0.4023916$. But this is not a flaw of our result, since with $\epsilon=0.999$, the column
algorithm does not stop until the second loop. After that, it includes
both the second and the third column, for which we can clearly see
$\boldsymbol{D}_{(2)}=\boldsymbol{D}_{(0)}+\boldsymbol{D}(2)+\boldsymbol{D}(3)$
and $\cos\left(\boldsymbol{D_{X}},\boldsymbol{D}_{(2)}\right)=1>0.999,\boldsymbol{D}_{(2)}=\boldsymbol{D_{X}}$.
Since the constant $C$ is generally not sharp, there is not a deterministic
relation between the choice of threshold $\epsilon$ and maximal error
between squared distances. However, if we want to control the error bound,
then the theorem will help us choosing $\epsilon$ given an $m\times p$
data matrix $\boldsymbol{X}$.

\pagebreak
\bibliographystyle{chicago}
\bibliography{bibliography}

\end{document}